\documentclass[12pt]{article}
\usepackage[dvips]{color}
\usepackage{graphics}
\usepackage[dvips]{graphicx}
\usepackage{mathrsfs}
\usepackage{amssymb}
\usepackage{verbatim}
\usepackage{float}
\newcommand{\be}{\begin{equation}}
\newcommand{\ee}{\end{equation}}
\newcommand{\bea}{\begin{eqnarray}}
\newcommand{\eea}{\end{eqnarray}}

\def\4vol{{\int d^4x \sqrt{-g}}}

\newcommand{\nc}{\newcommand}

\nc{\nt}{\tilde{N}}
\nc{\ra}{\rightarrow}
\nc{\lsim}{\begin{array}{c}\,\sim\vspace{-21pt}\\< \end{array}}
\nc{\gsim}{\begin{array}{c}\sim\vspace{-21pt}\\> \end{array}}
\nc{\tnt}{\tilde{N}}
\nc{\tst}{\tilde{t}}

\nc{\LL}{L}
\nc{\vv}{\tilde{v}}

\title{
\vspace*{-1.3cm}
\begin{flushright}
\normalsize{
ANL-HEP-PR-07-19\\
EFI-07-07\\
FERMILAB-PUB-07-074-T
}
\end{flushright}
\vspace{0.5cm}
\Large
\textbf{Challenges for MSSM Higgs searches at Hadron Colliders}
\vspace*{0.5cm}
\author{\textbf{M.~Carena$^a$, A.~Menon$^{b,c}$ and C.E.M.~Wagner$^{b,c,d}$}\\ 
\\[0.5cm]
$^a$\normalsize\emph{Theoretical Physics Dept., Fermi National Laboratory,
Batavia, IL 60510} \\
$^b$\normalsize\emph{HEP Division, Argonne National Laboratory,
9700 Cass Ave.,
Argonne, IL 60439, USA} \\
$^c$\normalsize\emph{Enrico Fermi Inst., Univ. of Chicago,
5640 S. Ellis Ave., Chicago, IL 60637, USA} \\
$^d$\normalsize\emph{KICP and Dept. of Physics, Univ. of Chicago,
5640 S. Ellis Ave.,Chicago IL 60637, USA}}}

\begin{document}
\setcounter{page}{0}
\maketitle
\vspace{0.5cm}
\begin{abstract}
In this article we analyze the impact of B-physics and Higgs physics
at LEP on standard and non-standard Higgs bosons searches 
at the Tevatron and the LHC, 
within the framework of minimal flavor violating 
supersymmetric models.
The B-physics constraints we consider come from the experimental
measurements of the
rare B-decays $b \to s \gamma$ and $B_u \to \tau \nu$  and the experimental 
limit on the $B_s \to \mu^+\mu^-$ branching ratio. We show that these 
constraints are severe for large values of the trilinear soft 
breaking parameter $A_t$, rendering the non-standard 
Higgs searches at hadron colliders less promising. On the contrary these
bounds are relaxed for small values of $A_t$ and large values of the Higgsino 
mass parameter $\mu$, enhancing the prospects for the direct detection
of non-standard Higgs bosons at both colliders.
We also consider the available ATLAS and CMS 
projected sensitivities in the standard model Higgs search channels, 
and we discuss the LHC's ability in probing the whole MSSM parameter space.
In addition we also consider the expected Tevatron
collider sensitivities in the standard model Higgs $h \to b \bar{b}$
channel to show that it may be able to find 3~$\sigma$ evidence in
the B-physics allowed regions for small or moderate values of the 
stop mixing parameter.
\end{abstract}
\thispagestyle{empty}

\pagebreak

\section{Introduction}

Over the last twenty years, the Standard Model (SM) has provided
an exceptionally accurate description of all high energy physics experiments
-- whether they be electroweak precision or flavor physics observables. 
The only part of the Standard Model that remains to be tested is the mechanism
for electroweak symmetry breaking. In the Standard Model, electroweak
symmetry breaking is achieved by the scalar Higgs field acquiring a vacuum 
expectation value (vev), thereby giving mass to the quarks, leptons and gauge 
bosons. However, this mechanism for electroweak symmetry breaking has a problem
in that the Higgs potential is unstable with respect to radiative corrections,
that is the scalar Higgs mass gets radiative corrections proportional to the 
cutoff due to fermion and boson loops. A number of extensions of the Standard 
Model have been suggested to try to alleviate this problem. Supersymmetry is 
one of the most promising of these extensions of the SM, in which every SM 
fermion (boson) has a spin-0 (spin-1/2) super-partner.

The minimal supersymmetric extension of the Standard Model or MSSM, with 
gauge invariant SUSY breaking masses of the order of 1 TeV, predicts an 
extended Higgs sector with a light SM-like Higgs boson of mass lower than about
130 GeV~\cite{mHiggsrad}--\cite{mhiggsEP5} that is in good agreement with
precision electroweak measurements. However the flavor structure of these SUSY 
breaking masses is not well understood. If there are no tree-level flavor 
changing neutral currents associated with the gauge and super-gauge 
interactions, the deviations from SM predictions are small. Such small 
deviations can be achieved if the quark and squark mass matrices are block 
diagonalizable in the same basis (an example is flavor blind squark and slepton
masses). The flavor violating effects in these minimal flavor violating models 
are induced by loop factors proportional to CKM matrix elements as in the 
Standard Model. The B-physics properties of these kinds of supersymmetric 
extensions of the SM have been studied in great detail in 
Refs.~\cite{Bertolini:1990if}--\cite{Foster:2005wb}.

The recent improvements in our understanding of B-physics observables have put 
interesting constraints on Higgs searches in the MSSM at the Tevatron and LHC
colliders. In Ref.~\cite{Carena:2006ai} we analyzed the constraints that the
non-observation of the $B_s \to \mu^+ \mu^-$ rare decay and the 
measurement of the $b \to s \gamma$ rare decay put on 
non-standard model Higgs searches at hadron colliders. In 
this article, we additionally explore  the regions of SUSY parameter space
that can be probed in SM-like Higgs searches for different benchmark scenarios.
We also extend our analysis in the B-physics sector to include the
additional information coming from the recent
measurement of $\mathcal{BR}(B_u \to \tau \nu)$ at Belle~\cite{Ikado:2006un} 
and Babar~\cite{Aubert:2006fk}. We find an interesting region of parameter 
space (i.e. large values of the Higgsino mass parameter $\mu$ and moderate 
values of the stop mixing parameter $X_t$) for which non-standard Higgs 
searches are not strongly constrained by B-physics. In particular, we find that
scenarios with small stop mixing, like
the so called minimal mixing scenario~\cite{Carena:2005ek}, and large 
Higgsino parameter $\mu$ look very promising for the Tevatron and the 
LHC. 
B-physics constraints in these scenarios seem to allow the region around a 
CP-odd Higgs mass $M_A \sim 160$~GeV and $\tan \beta \sim 50$ (where $\tan 
\beta = v_2/v_1$ is the ratio of the two Higgs vev's), which can be easily 
probed at the  Tevatron in the near future. For non-standard Higgs searches we 
show the present
D0~\cite{D01fbhtautau} and CDF~\cite{cdf1fbhtautau}
excluded regions in the $M_A - \tan \beta$ plane with 1 fb~$^{-1}$ of
data in the $\tau \tau$ inclusive channel and the Tevatron and LHC 
available projections for 4 fb$^{-1}$ and 30
fb$^{-1}$~\cite{TeVBstomumu,nikitenko_cms} respectively, that depend only 
slightly on the other low energy SUSY parameters. 
Small to moderate MSSM Higgs masses are also
interesting from the point of view of direct dark matter detection experiments,
since in that case t-channel Higgs exchange contributes importantly to 
neutralino dark matter scattering off nuclei. This contribution implies a 
strong connection between the constraints on SUSY parameters from direct dark
matter searches and non-standard MSSM Higgs searches at colliders.
In particular, the present direct detection limits 
on neutralino dark matter within the MSSM puts strong constraints on Higgs
searches unless the Higgsino component of the neutralino is quite small (i.e. 
large values of $\mu$), 
independent of the stop sector parameters~\cite{Carena:2006nv}.

This article is organized as follows. In section 2, we define our
theoretical setup for both the B-physics constraints and Higgs searches
within the MSSM. In section 3, we discuss representative benchmark
scenarios that have different properties for B-physics and Higgs
searches. We show that within the MSSM there is a strong
complementarity between the constraints coming from
non-standard Higgs searches and rare B-decays. Taking into
account these constraints we study the potential for standard model like
Higgs boson discovery at the Tevatron and the 
LHC~\cite{TeVBstomumu, nikitenko_cms}. 
For the Tevatron Higgs 
searches we assumed, conservatively, a final 
Tevatron luminosity of 4~fb$^{-1}$, while for Higgs searches at LHC, in the 
early phase, we used the expected 30~fb$^{-1}$ luminosity estimates. 
Finally we conclude in section~4.

\section{Theoretical Setup}

\subsection{Higgs Searches and Benchmark Scenarios} 

\subsubsection{Couplings and Masses of the Higgs Sector in the MSSM}

In the MSSM there are three neutral scalar Higgs fields. Assuming no extra 
sources of CP violation in the MSSM beyond that of the SM, there are two 
CP-even Higgs 
bosons which are admixtures of the real neutral $H_1^0$ and $H_2^0$ components
\begin{eqnarray}
\left(\begin{array}{c}
h \\
H
\end{array} \right) = 
\left(\begin{array}{cc}
-\sin \alpha & \cos \alpha \\
\cos \alpha & \sin \alpha
\end{array} \right)
\left(\begin{array}{c}
H_1^0 \\
H_2^0
\end{array} \right)
\end{eqnarray}
and an additional CP-odd Higgs field $A$, where $\alpha$ is the mixing angle 
that diagonalizes the CP-even Higgs mass matrix. The tree-level Higgs couplings
to the SM fermions and gauge bosons are given 
by~\cite{CarenaMrenna,Roszkowski:2006mi}
\begin{eqnarray}
\frac{1}{(\phi d \bar{d})_{SM} \left((\phi u \bar{u})_{SM}\right)} \left(
\begin{array}{c}
(h d \bar{d})_{MSSM} \left((h u \bar{u})_{MSSM}\right) \\
(H d \bar{d})_{MSSM} \left((H u \bar{u})_{MSSM}\right)\\
(A d \bar{d})_{MSSM} \left((A u \bar{u})_{MSSM}\right)
\end{array} \right)
&=& \left( 
\begin{array}{c}
-\sin \alpha/\cos \beta \;\; (\cos \alpha/\sin \beta) \\
\cos \alpha/\cos \beta \;\; (\sin \alpha/\sin \beta)\\
\tan \beta \;\; (\cot \beta)
\end{array} \right) \nonumber \\
\frac{1}{(\phi V V)_{SM}}\left(
\begin{array}{c}
(h V V)_{MSSM} \\
(H V V)_{MSSM} \\
(A V V)_{MSSM}
\end{array}\right)
&=& \left(
\begin{array}{c}
\sin (\beta-\alpha) \\
\cos (\beta-\alpha) \\
0
\end{array}\right) \label{coupl:eq}
\end{eqnarray}
where $V$ can be either the $Z$ or $W$ vector boson. At moderate or large 
values of $\tan \beta$, one of the two CP-even Higgs bosons tends to couple 
strongly to the gauge bosons while the other one only couples weakly. We will 
denote the Higgs boson that couples to the gauge bosons the strongest as 
SM-like. The CP-odd and the other CP-even Higgs bosons are
denoted as non-standard and have $\tan \beta$ enhanced couplings to the down
quarks and leptons (see Eq.~\ref{coupl:eq}).

The identification of the SM-like Higgs depends critically on the size of the 
pole mass of the pseudo-scalar Higgs $M_A$. For large values of $M_A$, the 
lighter Higgs becomes SM-like and its mass  has the approximate analytic 
form~\cite{mHiggsrad,mhiggsRG1,Carena:1995wu}
\begin{eqnarray}
(M_h^{max})^2 &=& M_Z^2 \cos^2 (2\beta) (1 - \frac{3m_t^2}{8\pi^2v^2} t) 
\nonumber\\
&+& \frac{3m_t^4}{4\pi^2v^2} \left[\frac{1}{2} \tilde{X}_t + t + 
\frac{1}{16\pi^2} \left(\frac{3 m_t^2}{2v^2} - 32 \pi \alpha_3\right)(
\tilde{X}_t t + t^2)\right], \label{mhmax:eq}
\end{eqnarray}
where $\tilde{X}_t =  \frac{2X_t^2}{M_{SUSY}^2} - \frac{X_t^4}{6M_{SUSY}^4}$,
$X_t = A_t -\mu/\tan \beta$,
$t = \log\left(\frac{M_{SUSY}^2}{m_t^2}\right)$ and $M_{SUSY}$ is the geometric
mean of the stop masses. In Eq.~(\ref{mhmax:eq}), we have included the leading
two-loop radiative corrections from the stop sector but we have not included 
the two-loop corrections associated with the relation between the top quark
mass and the top Yukawa coupling at the stop mass scale, that depends on the 
relative sign of the gluino mass and $X_t$~\cite{bse}.
At values of the CP-odd Higgs boson mass $M_A$ less than $m_h^{max}$ and 
large values of $\tan \beta$, $\alpha \sim \beta$ and the heavier CP-even Higgs
is SM-like with mass given approximately by Eq.~(\ref{mhmax:eq}). 

\subsubsection{SM-like Higgs Boson Searches}

The CMS and ATLAS collaborations have calculated the signal significance curves
for standard model Higgs detection at the LHC. Due to the modified Higgs 
couplings in the MSSM, for the same Higgs masses, these estimates can change 
significantly with changes in the supersymmetric mass parameters. To quantify 
when the significance will be either enhanced or reduced we consider the 
quantity~\cite{CarenaMrenna,Roszkowski:2006mi}
\begin{eqnarray}
R = \frac{\sigma (P \bar{P} \rightarrow X \phi)_{MSSM} \mathcal{BR}(\phi 
\rightarrow Y)_{MSSM}}{\sigma (P \bar{P} \rightarrow X \phi)_{SM} 
\mathcal{BR}(\phi \rightarrow Y)_{SM}} \label{rratio:eq}
\end{eqnarray} 
where $X$ are particles produced in association with the Higgs and $Y$
are SM decay products of the Higgs\footnote{For the region of parameter space 
we study only standard model decays are open.}. As the predicted SM-like Higgs 
mass range 
within the MSSM is less than or about $130$~GeV, we only consider the light 
Higgs production and decay channels $q \bar{q} \phi \rightarrow q \bar{q} \tau 
\bar{\tau}$ and $\phi \rightarrow \gamma \gamma$ at the LHC and $W/Z \phi
(\phi \to b \bar{b}) $ at the Tevatron. At a luminosity larger than
30~fb$^{-1}$ at the LHC, the $t\bar{t} \phi$ will become effective. However as 
we are considering only the early phase of the LHC we will not study this 
process.

For the $q \bar{q} \phi \rightarrow q \bar{q} \tau \bar{\tau}$ channel the 
Higgs is produced dominantly by weak-boson fusion. Hence, the tree-level 
production cross-section is proportional to the square of the $(\phi VV)_{SM}$ 
coupling in 
Eq.~(\ref{coupl:eq}), which implies that the ratio of production cross-sections
in Eq.~(\ref{rratio:eq}) is proportional to $\sin^2 (\beta -\alpha) (\cos^2 
(\beta -\alpha))$ when $M_A$ is larger (smaller) than $M_h^{max}$.
At large $\tan \beta$ and $M_A > M_h^{max}$ ($M_A < M_h^{max}$) the 
Higgs mixing angle $\sin \alpha \sim -1/\tan \beta$ ($\cos \alpha \sim 1/\tan 
\beta$). Hence, in this region of the $M_A \-- \tan \beta$ plane the 
$(hVV)_{MSSM}$ ($(HVV)_{MSSM}$) couplings are very close to their SM values. 
Therefore at large $\tan \beta$ and small or 
large values of $M_A$, compared to $M_h^{max}$, the ratio $\sigma 
(P \bar{P} \rightarrow X \phi)_{MSSM}/\sigma (P \bar{P} \rightarrow X \phi)_{
SM}$ is close to one. For $\phi \rightarrow \gamma 
\gamma$ channel the Higgs is mainly produced through gluon fusion which is 
induced by third generation quark and squark loops. For squark masses greater
than $500$~GeV, like those we are considering in this paper, the squark 
contributions are small and the SM-like Higgs has a production 
cross-section similar to that of the standard model Higgs.

Whenever $M_A$ is comparable to the SM-like Higgs mass, $|M_A -m_{h}^{max}| 
\lsim 10$~GeV, both the CP-even Higgs bosons acquire similar masses and 
have non-standard gauge and yukawa couplings.
Hence for each of these channels we follow the prescription given in
Ref.~\cite{CarenaMrenna} and sum the contributions from both
the CP-even Higgs states so that
\begin{eqnarray}
R = \frac{\sigma (P \bar{P} \rightarrow X h)_{MSSM} \mathcal{BR}(h 
\rightarrow Y)_{MSSM} + \sigma (P \bar{P} \rightarrow X H)_{MSSM} 
\mathcal{BR}(H \rightarrow Y)_{MSSM}}{\sigma (P \bar{P} \rightarrow X 
\phi)_{SM} \mathcal{BR}(\phi \rightarrow Y)_{SM}},
\end{eqnarray}
because we assume that the two signals cannot be separated.

If $M_A$ is larger (smaller) than $M_h^{max}$ and the loop corrections
to the off-diagonal elements of the CP-even Higgs mass matrix are small, then 
the large $\tan \beta$ induced corrections do not enhance or reduce the 
$h b \bar{b}$ ($H b \bar{b}$) or $h \tau \bar{\tau}$ ($H \tau \bar{\tau}$) 
couplings and they remain Standard Model like. Hence, in these regions of 
parameter space the branching ratios into either $b$'s or $\tau$'s are close to
their Standard Model values. The $\phi \gamma \gamma$ coupling is induced 
through quark loops and hence is generally small. However, in scenarios where 
the $\phi b \bar{b}$ and $\phi \tau \bar{\tau}$ couplings are suppressed, like
for example if there is a cancellation of the off-diagonal CP-even mass Higgs 
matrix element due to radiative effects, the $\phi \to \gamma \gamma$ branching
ratio can be relatively enhanced. We shall discuss this case in 
section~\ref{smallalpeff:sec}.

\subsubsection{Non-standard Higgs Boson Searches} 

At large $\tan \beta$ the non-standard Higgs bosons are produced in 
association with bottom quarks or through gluon fusion. For both of these 
processes, at large $\tan \beta$, the relevant coupling is the bottom 
Yukawa coupling~\cite{Carena:2005ek,Aglietti:2006ne}. Therefore including the 
relevant large $\tan \beta$ radiative correction we find the production 
cross-section is proportional to the square of the bottom Yukawa 
$y_b^2 = (y_b^{SM})^2 \tan^2 \beta/(1+\epsilon_3 \tan \beta)^2
$, where the precise definition of this loop induced correction is given in 
Eq.~(\ref{epsilon3:eq}). In addition, at large $\tan \beta$~\cite{Carena:2005ek,
Aglietti:2006ne} the branching ratio of the decay of the non-standard Higgs 
boson into $\tau \tau$ is approximately given by  
\begin{eqnarray}
Br(A,H \to \tau^+ \tau^-) \simeq \frac{(1+\epsilon_3 \tan \beta)^2}{(1
+\epsilon_3 \tan \beta)^2 + 9}.
\end{eqnarray}
Hence the total production rate of the CP-odd Higgs boson at large $\tan \beta$
is
\begin{eqnarray}
\sigma(gg,b\bar{b} \rightarrow A) \times \mathcal{BR}(A\rightarrow \tau^+ 
\tau^-) \sim \sigma(gg,b \bar{b} \rightarrow A)_{SM} \frac{\tan^2 \beta}{(1+
\epsilon_3 \tan \beta)^2 + 9} \label{nonsmhprod:eq}.
\end{eqnarray}
Therefore we can define a ratio similar to Eq.~(\ref{rratio:eq}) 
\begin{eqnarray}
r = \frac{\sigma (gg,b \bar{b} \rightarrow A)_{MSSM} \mathcal{BR}(A 
\rightarrow \tau^+ \tau^-)_{MSSM}}{\sigma (gg,b \bar{b}\rightarrow \phi)_{SM}
\mathcal{BR}(\phi \rightarrow \tau^+ \tau^-)_{SM}} \sim \frac{\tan^2 \beta}{(1+
\epsilon_3 \tan \beta)^2 + 9} \label{rratio_nsmh:eq}
\end{eqnarray}
and a analogous expression holds for the CP-even non-standard Higgs boson 
production and decay rates.

\subsection{B Physics Observables and Limits} 

We will consider the four B physics observables: $\mathcal{BR}(B_s \to \mu^+ 
\mu^-)$, $\Delta M_s$, $\mathcal{BR}(b \to s \gamma)$ and $\mathcal{BR}(
B_u \to \tau \nu)$ within the minimal flavor violating MSSM.

\subsubsection{$\mathcal{BR}(B_s \to \mu^+ \mu^-)$}

In the Standard Model the relevant contribution to the $B_s \to \mu^+ \mu^-$ 
process comes through the Z-penguin and the W-box diagrams which have the 
analytic form~\cite{Dedes:2002er,Buchalla:1995vs}
\begin{eqnarray}
\mathcal{BR}(B_s \to \mu^+ \mu^-)_{SM} = \frac{G_F^2 \alpha_{em}^2}{16 \pi^3} 
M_{B_s} \tau_{B_s} F_{B_s}^2 |V_{tb} V_{ts}|^2 \sqrt{1-\frac{4m_{\mu}^2}{
M_{B_s}^2}} m_{\mu}^2 C_{10}^2(x_t)
\end{eqnarray}
where $\tau_{B_s}$ is the mean lifetime, $F_{B_s}$ is the decay constant of the
$B_s$ meson, $x_t = m_t/M_W$ and
\begin{eqnarray}
& & C_{10}(x) = b_0(x) - c_0(x) \\
& & c_0(x) = \frac{x}{8} \left[\frac{x-6}{x-1} + \frac{3x+2}{(x-1)^2} \ln(x) 
\right] \\
& & b_0(x) = \frac{1}{4} \left[\frac{x}{1-x} + \frac{x}{(x-1)^2} \ln(x) 
\right].
\end{eqnarray}
Therefore the predicted SM value comes out to be~\cite{Dedes:2002er,
Buchalla:1995vs}
\begin{eqnarray}
\mathcal{BR}(B_s \to \mu^+ \mu^-)_{SM} = (3.8 \pm 0.1) \times 10^{-9}.
\end{eqnarray}
However in the presence of supersymmetry at large $\tan \beta$, there are 
significant contributions from Higgs mediated neutral currents, which have the 
form~\cite{Buras:2002wq,Buras:2002vd}
\begin{eqnarray}
\mathcal{BR}(B_s \to \mu^+ \mu^-) &=& 3.5 \times 10^{-5} 
\left[\frac{\tan \beta}{50} \right]^6 
\left[\frac{\tau_{B_s}}{1.5 {\rm ps}}\right]
\left[\frac{F_{B_s}}{230 {\rm MeV}}\right]^2
\left[\frac{|V_{ts}|}{0.040}\right]^2 \nonumber \\
&\times& \frac{m_t^4}{M_A^4} \frac{(16 \pi^2 \epsilon_Y)^2}{(1+\epsilon_3 \tan 
\beta)^2(1+\epsilon_0 \tan \beta)^2} \label{BRbsmumu:eq}
\end{eqnarray}
where 
\begin{eqnarray}
\epsilon_3 = \epsilon_0 + y_t^2 \epsilon_Y \label{epsilon3:eq}. 
\end{eqnarray}
The gluino loop factor 
$\epsilon_0$ and the chargino-stop loop factor $\epsilon_Y$ are given by
\begin{eqnarray}
\epsilon_0 \approx \frac{2 \alpha_s}{3\pi} M_3 \mu C_0(m_{\tilde{b}_1}^2,
m_{\tilde{b}_2}^2,M_3^2)\\ 
\epsilon_Y \approx \frac{1}{16\pi^2} A_t \mu C_0(m_{\tilde{t}_1}^2,
m_{\tilde{t}_2}^2,\mu^2)
\end{eqnarray}
respectively, where $m_{\tilde{b}_i}$ is the i$^{th}$ sbottom mass, 
$m_{\tilde{t}_i}$ is the i$^{th}$ stop mass, $M_3$ is the gluino mass, 
$\mu$ is the higgsino mass parameter, $A_t$ is the soft SUSY breaking stop
trilinear parameter and  
\begin{eqnarray}
C_0(x,y,z) = \frac{y}{(x-y)(z-y)} \log(y/x) + \frac{z}{(x-z)(y-z)} \log(z/x).
\end{eqnarray}

The present experimental exclusion limit at 95\% C.L. from 
CDF~\cite{Bernhard:2005yn} is
\begin{eqnarray}
\mathcal{BR}(B_s\rightarrow \mu^+ \mu^-) \leq 1 \times 10^{-7},
\label{bsmumubound:eq}
\end{eqnarray}
which puts strong restrictions on possible flavor changing neutral currents in 
the MSSM at large $\tan\beta$. Additionally, if no signal is observed, the 
projected exclusion limit, at 95\% C.L., on this process for 4~fb$^{-1}$ at the
Tevatron is~\cite{TeVBstomumu} 
\begin{eqnarray}
\mathcal{BR}(B_s\rightarrow \mu^+ \mu^-) \leq 2.8 \times 10^{-8}.
\label{bsmumuboundf:eq}
\end{eqnarray}
Similarly, if no signal is observed at the LHC, the projected ATLAS bound at 
10~fb$^{-1}$ is~\cite{LHCBstomumu}
\begin{eqnarray}
\mathcal{BR}(B_s\rightarrow \mu^+ \mu^-) \leq 5.5 \times 10^{-9}. 
\label{bsmumuboundLHC:eq}
\end{eqnarray}
Therefore considering Eq.~(\ref{BRbsmumu:eq}) in the absence of a signal, 
these experiments will put very strong constraints on the allowed MSSM 
parameter space. In addition, LHCb has the potential to claim a $3 \sigma \; (
5 \sigma)$ evidence (discovery) of a standard model signature with as little
as $\sim 2$fb$^{-1}$($6$fb$^{-1}$) of data~\cite{LHCB}.
 
\subsubsection{$\Delta M_s$}

In the Standard Model the dominant contribution to $\Delta M_s$ comes from 
W-top box diagrams that have the analytical 
form~\cite{Buras:2002wq,Buras:2002vd}
\begin{eqnarray}
\Delta M_s = \frac{G_F^2 M_W^2}{6\pi^2} M_{B_s} \eta_2 F_{B_s}^2 \hat{B}_{B_s}
|V_{ts}|^2 S_0(m_t)
\end{eqnarray}
where $M_{B_s}$ is the $B_s$ meson mass, $\hat{B}_{B_s}$ is the $B_s$ bag 
parameter, 
$\eta_2$ is the NLO QCD factor and
\begin{eqnarray}
S_0(m_t) \simeq 2.39 \left(\frac{m_t}{167 \mbox{GeV}}\right)^{1.52}.
\end{eqnarray}
The updated theoretical predictions from the CKMfitter and UTFit
groups are slightly different. The UTFit group finds the 95 \% C.L. 
range~\cite{Bona:2006ah}
\begin{eqnarray}
(\Delta M_s)^{SM} = (20.9 \pm 2.6) {\rm ps}^{-1} 
\label{deltamsbound:eq}
\end{eqnarray}
which is consistent with the 
CKMfitter groups' $2\sigma$ range~\cite{ckmfitter}
\begin{eqnarray}
13.4 \; {\rm ps}^{-1} \leq (\Delta M_s)^{SM} \leq 31.1 \; {\rm ps}^{-1} 
\label{deltamsbounda:eq}
\end{eqnarray}
and central value of $18.9 \; {\rm ps}^{-1}$.

About a year ago, the D0 collaboration reported a signal consistent with 
values of $\Delta M_s$ in the range
\begin{eqnarray}
21 \; (\mbox{ps})^{-1} > \Delta M_s > 17 \; \mbox{ps}^{-1}
\end{eqnarray} 
at the 90 \% C.L.~\cite{Abazov:2006dm}. More recently, the CDF 
collaboration has made a measurement, with the result~\cite{Giagu:2006qv}  
\begin{eqnarray}
\Delta M_s = (17.77 \pm 0.10(\mbox{stat})\pm0.07(\mbox{syst})) \mbox{ps}^{-1}.
\end{eqnarray}

The large theoretical uncertainties and the precise 
experimental value suggest that small or moderate negative contributions to 
$\Delta M_s$ may be easily accommodated. As shown in 
Refs.~\cite{Isidori:2001fv,Buras:2002wq,Buras:2002vd,Dedes:2002er} for large 
$\tan \beta$ and uniform squark masses one obtains negative contributions to
$\Delta M_s$ that are well approximated by
\begin{eqnarray}
(\Delta M_s)^{DP} &=& - 12.0 {\rm ps}^{-1} \left[\frac{\tan \beta}{50}\right]^4
\left[\frac{F_{B_s}}{230 MeV}\right]^2 \left[\frac{V_{ts}}{0.04}\right]^2 
\nonumber \\
& & [\frac{\bar{m}_b(\mu_s)}{3.0 GeV}] [\frac{\bar{m}_s(\mu_s)}{0.06 GeV}] 
[\frac{\bar{m}_t^4(\mu_s)}{M_W^2 M_A^2}] \frac{(16 \pi^2 \epsilon_Y^2)^2}{(1+
\epsilon_3 \tan \beta)^2(1+\epsilon_0 \tan \beta)}. \label{dmsdp:eq}
\end{eqnarray}
In the next section we will discuss the interplay between the 
$\mathcal{BR}(B_s \to \mu^+ \mu^-)$ in Eq.~(\ref{BRbsmumu:eq}) and 
$\Delta M_s$ in Eq.~(\ref{dmsdp:eq}) within the framework of minimal flavor
violating MSSM.

\subsubsection{$\mathcal{BR}(b \to s \gamma)$} \label{bsg:sec}

The next B-physics process of interest is the rare decay $b \to s \gamma$.
The world experimental average of the branching of this rare decay 
is~\cite{:2006bi,Becher:2006pu}
\begin{eqnarray}
\mathcal{BR}(b\to s \gamma)^{exp} = (3.55 \pm 0.24^{+0.09}_{-0.10} \pm 0.03) 
\times 10^{-4}.
\label{bsgbound:eq}
\end{eqnarray} 
This experimental result is close to the SM central value and so puts 
constraints on flavor violation in any extension of the Standard 
Model. However, the theoretical uncertainties 
in the Standard Model for this process are quite large~\cite{Becher:2006pu}
\begin{eqnarray}
\mathcal{BR}(b \to s \gamma)^{SM} = (2.98 \pm 0.26) \times 10^{-4}. 
\label{bsgsm:eq}
\end{eqnarray}
Using the experimental and SM ranges 
for the $\mathcal{BR}(b \to s \gamma)$ we find the $2\sigma$ allowed range is
\begin{eqnarray}
0.92 \leq   \frac{\mathcal{BR}(b \to s \gamma)^{MSSM}}{ BR(b \to s \gamma
)^{SM}} \leq 1.46.
\end{eqnarray}
This bound is appropriate for constraining new physics contributions due to 
the cancellation of the dominant uncertainties coming from infrared physics 
effects.

In minimal flavor violating MSSM there are two new contributions from the 
charged Higgs and the 
chargino-stops diagrams. The charged Higgs amplitude, including the stop 
induced two-loop effects, is proportional to the 
factor~\cite{Degrassi:2000qf,Carena:2000uj}
\begin{eqnarray}
A_{H+} \propto \left[\frac{1 - \frac{2 \alpha_s}{3\pi} \mu M_3 \tan \beta
\left(\cos^2 \theta_{\tilde{t}} C_0(m_{\tilde{s}_L}^2,m_{\tilde{t}_1}^2,M_3^2) 
+ \sin^2 \theta_{\tilde{t}} C_0(m_{\tilde{s}_L}^2,m_{\tilde{t}_2}^2,M_3^2) 
\right)}{1+\epsilon_3 \tan \beta}\right]\frac{m_t^2}{m_H^{+2}}, 
\label{bsgh+:eq}
\end{eqnarray}
where $\theta_{\tilde{t}}$ is the stop mixing angle.
The chargino-stop amplitude has the form~\cite{Degrassi:2000qf,Carena:2000uj}
\begin{eqnarray}
A_{\chi^-} \propto \frac{\mu A_t \tan \beta}{1+ \epsilon_3 \tan \beta} 
f(m_{\tilde{t}_1}^2,m_{\tilde{t}_2}^2,m_{\chi^-}^2). \label{bsgchi:eq}
\end{eqnarray}
where $f(m_{\tilde{t}_1}^2,m_{\tilde{t}_2}^2,m_{\tilde{\chi}^-}) \sim 
1/max(m_{\tilde{t}_1}^2,m_{\tilde{t}_2}^2)$ is the 
one-loop factor that depends on the stop masses and the chargino mass.
The specific dependences of these amplitudes on MSSM parameters are important 
in understanding the constraints on the SUSY contributions to $\mathcal{BR}(b 
\to s \gamma)$, which will be discussed below.

\subsubsection{$\mathcal{BR}(B_u \to \tau \nu)$}

The final B-physics observable of interest is the process 
$B_u \to \tau \nu$ which the Belle experimental collaboration
finds to be~\cite{Ikado:2006un}
\begin{eqnarray}
\mathcal{BR}(B_u \to \tau \nu)^{\rm Belle} = (1.79^{+0.56}_{-0.49}(\mbox{stat})^{
+0.46}_{-0.51}({\rm syst}))
\times 10^{-4},
\end{eqnarray}
while the Babar collaboration finds a value~\cite{Aubert:2006fk}
\begin{eqnarray}
\mathcal{BR}(B_u \to \tau \nu)^{\rm Babar} = (0.88^{+0.68}_{-0.67}(\mbox{stat})
\pm 0.11 ({\rm syst})) \times 10^{-4}.
\end{eqnarray}
The two values are within $2 \sigma$ of each other and 
both of them are consistent with the standard model prediction.
The average of these two experiments is~\cite{Bona:2006ah}
\begin{eqnarray}
\mathcal{BR}(B_u \to \tau \nu)^{\rm Exp} = (1.31 \pm 0.48 ) \times 10^{-4}.
\label{butaunuexp:eq}
\end{eqnarray}

The Standard Model contribution is mediated by the W-boson and has the generic 
form~\cite{Isidori:2006pk}
\begin{eqnarray}
\mathcal{BR}(B_u \to \tau \nu)^{\rm SM} = \frac{G_F^2 m_B m_{\tau}^2}{8\pi}
\left(1-\frac{m_{\tau}^2}{m^2_B} \right)^2 F_B^2 |V_{ub}|^2 \tau_B
\end{eqnarray}
and using the UTFit fitted value for $|V_{ub}| = (3.68 \pm 0.14) \times 10^{-3}$ 
(which is also in good agreement with the CKMfitter value~\cite{ckmfitter}), $\tau_B$ and the 
extracted value 
of $F_B = 0.237 \pm 0.037$~GeV leads to the value~\cite{Bona:2006ah} 
\begin{eqnarray}
\mathcal{BR}(B_u \to \tau \nu)^{\rm SM} = (0.85 \pm 0.13) \times 10^{-4}
\label{butaunu_sm:eq}. 
\end{eqnarray}
Observe, however that the value of $|V_{ub}| = (4.49 \pm 0.33) \times 10^{-9}$, 
extracted from inclusive semileptonic 
decays is higher and leads to the standard model prediction 
$\mathcal{BR}(B_u \to \tau \nu)^{\rm SM} = (1.39 \pm 0.44) \times 
10^{-4}$~\cite{Bona:2006ah}. 

\begin{figure}
\begin{center}
\resizebox{12.cm}{!}{\includegraphics{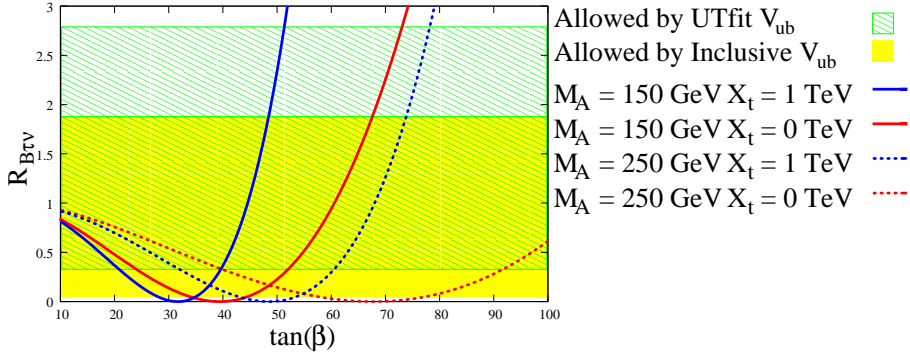}}
\end{center}
\caption{The green(grey) hatched area is the 2$\sigma$ allowed region of the 
ratio $R_{B\tau \nu}$ if the fitted value of $|V_{ub}|$ is used to calculate the 
standard model prediction of the $B_u \to \tau \nu$ decay rate. The 
yellow(light grey) is corresponding region if the inclusive determination of 
$|V_{ub}|$ is used instead of the fitted value. The solid (dashed) lines show the 
variation of $R_{B\tau \nu}$ with respect to $\tan \beta$ for 
$M_A=150{\rm GeV}(250{\rm GeV})$, while the red (grey) color and blue (dark 
grey) color correspond to $X_t=0$ and $X_t=1$~TeV respectively.
}
\label{rbtnu:fig}
\end{figure}

In the MSSM there is an extra contribution due to the charged Higgs which 
interferes destructively with the SM contribution, so that at large $\tan 
\beta$ the ratio of the two is~\cite{Isidori:2006pk,Hou:1992sy,Akeroyd:2003zr}
\begin{eqnarray}
R_{B\tau \nu} = \frac{\mathcal{BR}(B_u \to \tau \nu)^{\rm MSSM}}{\mathcal{BR}(
B_u \to \tau \nu)^{\rm SM}} = \left[1 - \left(\frac{m_B^2}{m_{H^{\pm}}^2} 
\right) \frac{\tan^2 \beta}{1+\epsilon_0 \tan \beta}\right]^2. 
\label{rbtaunu:eq}
\end{eqnarray}
Now assuming a 2$\sigma$ deviation in Eq.~(\ref{butaunuexp:eq}) 
and Eq.~(\ref{butaunu_sm:eq}) that is due to the charged Higgs 
contribution, we find the allowed range of values for this ratio to be 
\begin{eqnarray}
0.32 \leq R_{B\tau \nu} \leq 2.77. \label{btaunubound:eq}
\end{eqnarray}
However as discussed above, if the inclusive determination of $|V_{ub}|$ 
is used instead of the fitted value we get a different range of allowed values for
$R_{B\tau \nu}$. In Fig.~\ref{rbtnu:fig} we show the effect of choosing the 
$|V_{ub}|$ inclusive value over the fitted value. The green (grey) hatched 
region is allowed if we use the fitted value of $|V_{ub}|$ while the yellow (light
grey) region is allowed if we use the extracted value of 
$|V_{ub}|$ from inclusive semileptonic b-decays. From Fig.\ref{rbtnu:fig} we can see that if $M_A = 150$~GeV and 
$X_t=0$ the allowed values are $\tan \beta \sim 10\-- 25$ and $\tan \beta \sim 
53 \-- 70$ using the fitted value of $|V_{ub}|$, while using the inclusive 
value of $|V_{ub}|$ we find $ 10 \lsim \tan \beta \lsim 37$ or $43 \lsim \tan 
\beta \lsim 63$. Therefore, when we project this constraint onto the $M_A \-- 
\tan \beta$ plane the allowed regions are significantly different, 
especially at larger values of $M_A$. In particular the region of intermediate
$\tan \beta$ that is excluded by the $B_{u} \to \tau \nu$ constraint is much
smaller if we use the inclusive value of $|V_{ub}|$ instead of the fitted value
because the lower bound on $R_{B\tau \nu}$ is smaller for the value extract
from inclusive b-decays. Whenever we consider the constraint on the $B_u \to 
\tau \nu$ rate in this paper we will use the fitted values, so expect our bounds 
to be quite conservative and one could enlarge the B physics allowed region
by going to larger values of $|V_{ub}|$.

\section{B physics constraints and Higgs searches at hadron 
Colliders}

\begin{figure}
\begin{center}
\resizebox{7.cm}{!}{\includegraphics{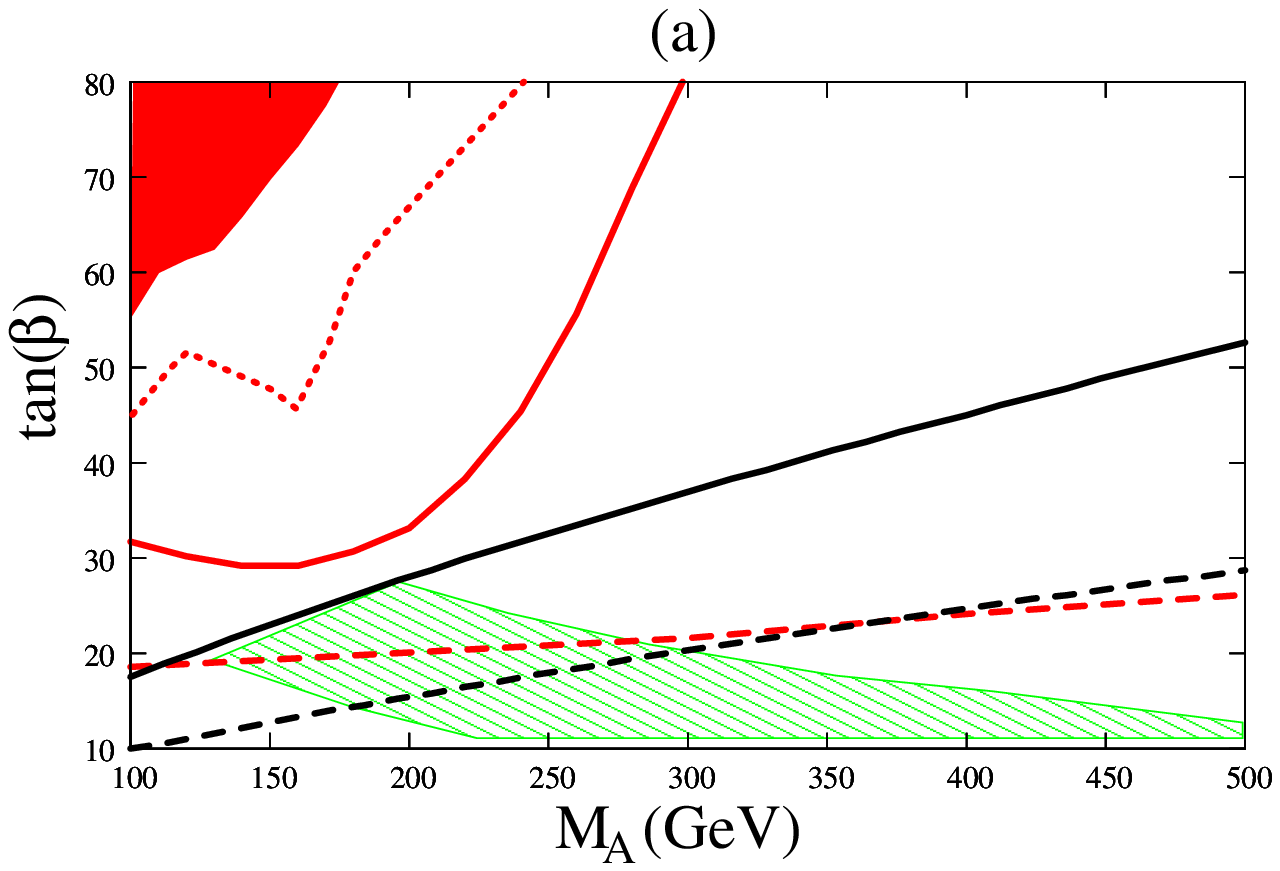}}
\resizebox{7.cm}{!}{\includegraphics{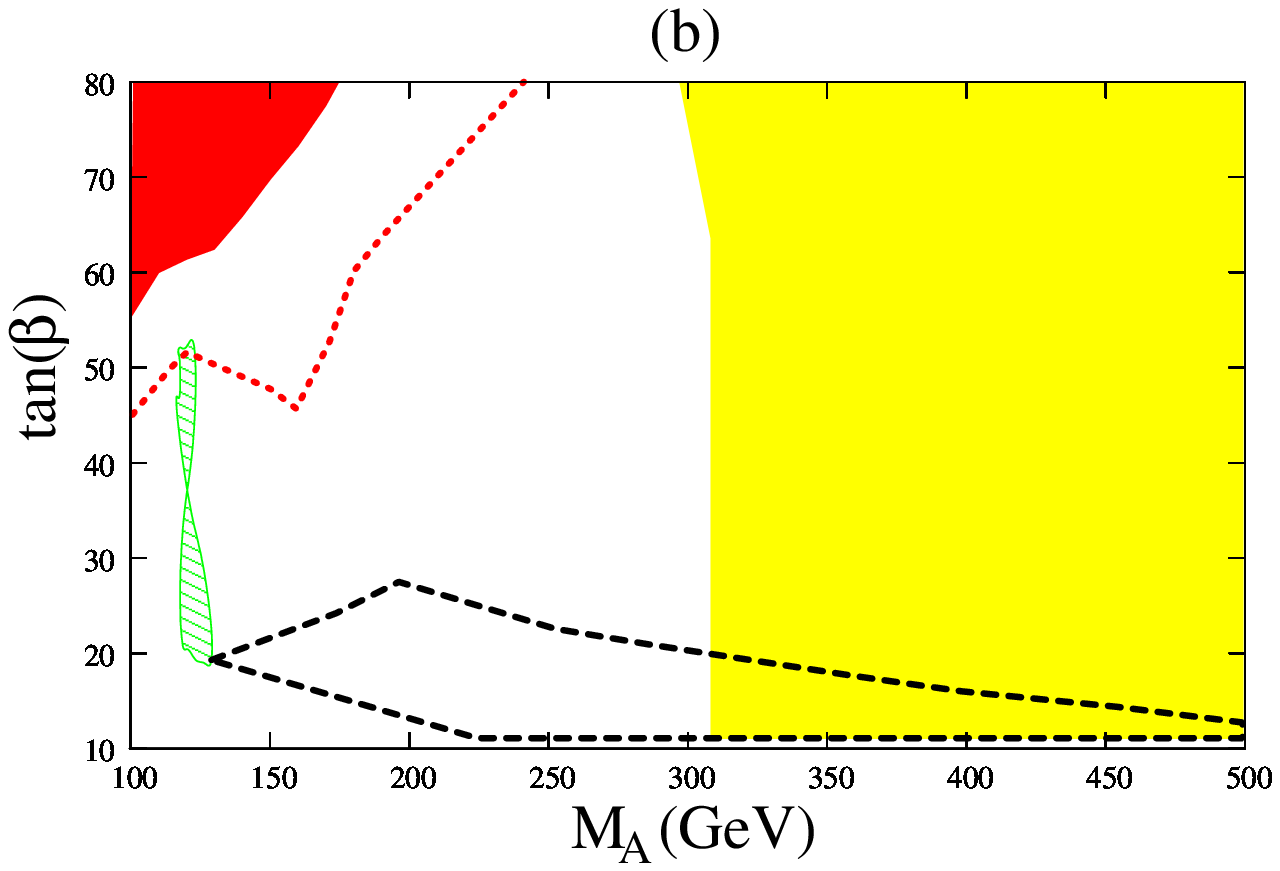}}
\resizebox{7.cm}{!}{\includegraphics{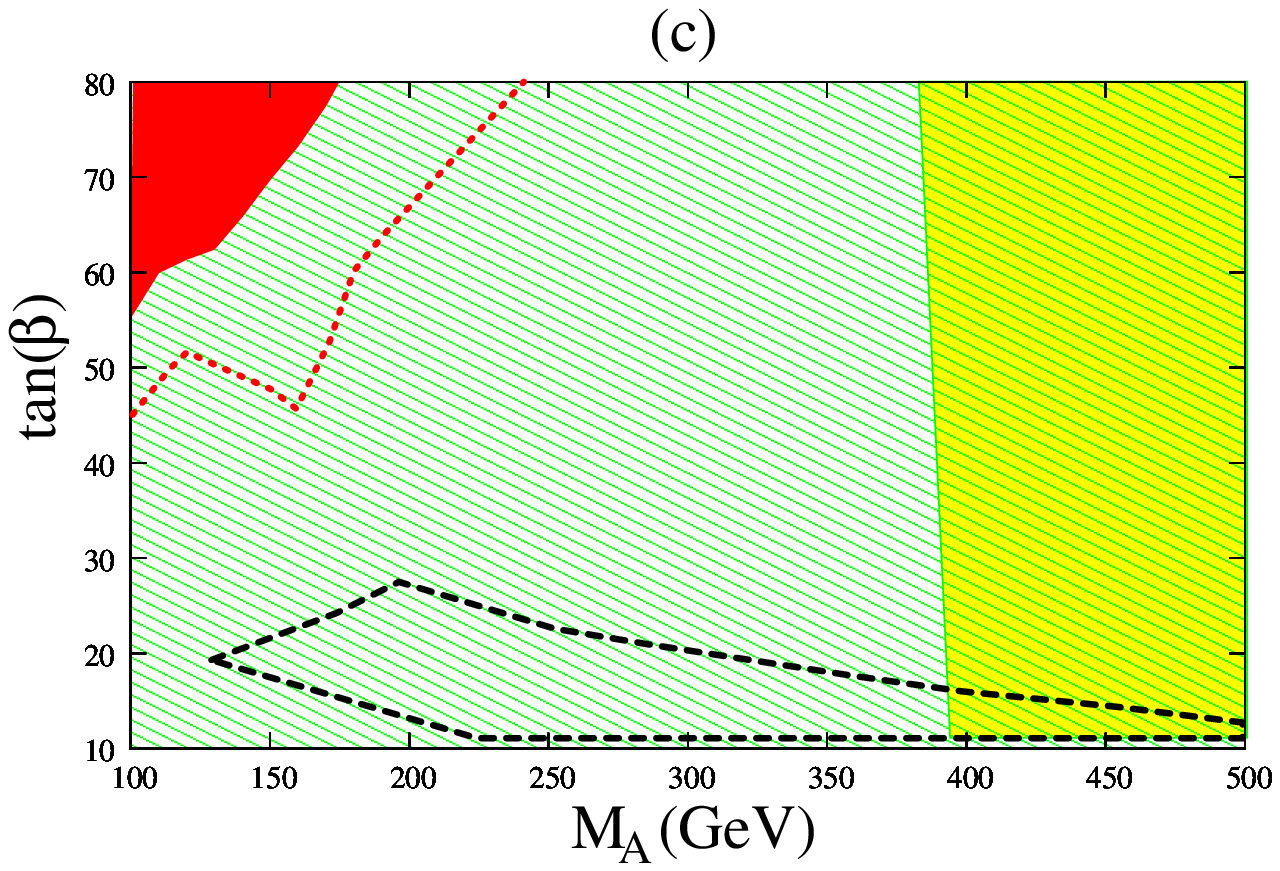}}
\resizebox{7.cm}{!}{\includegraphics{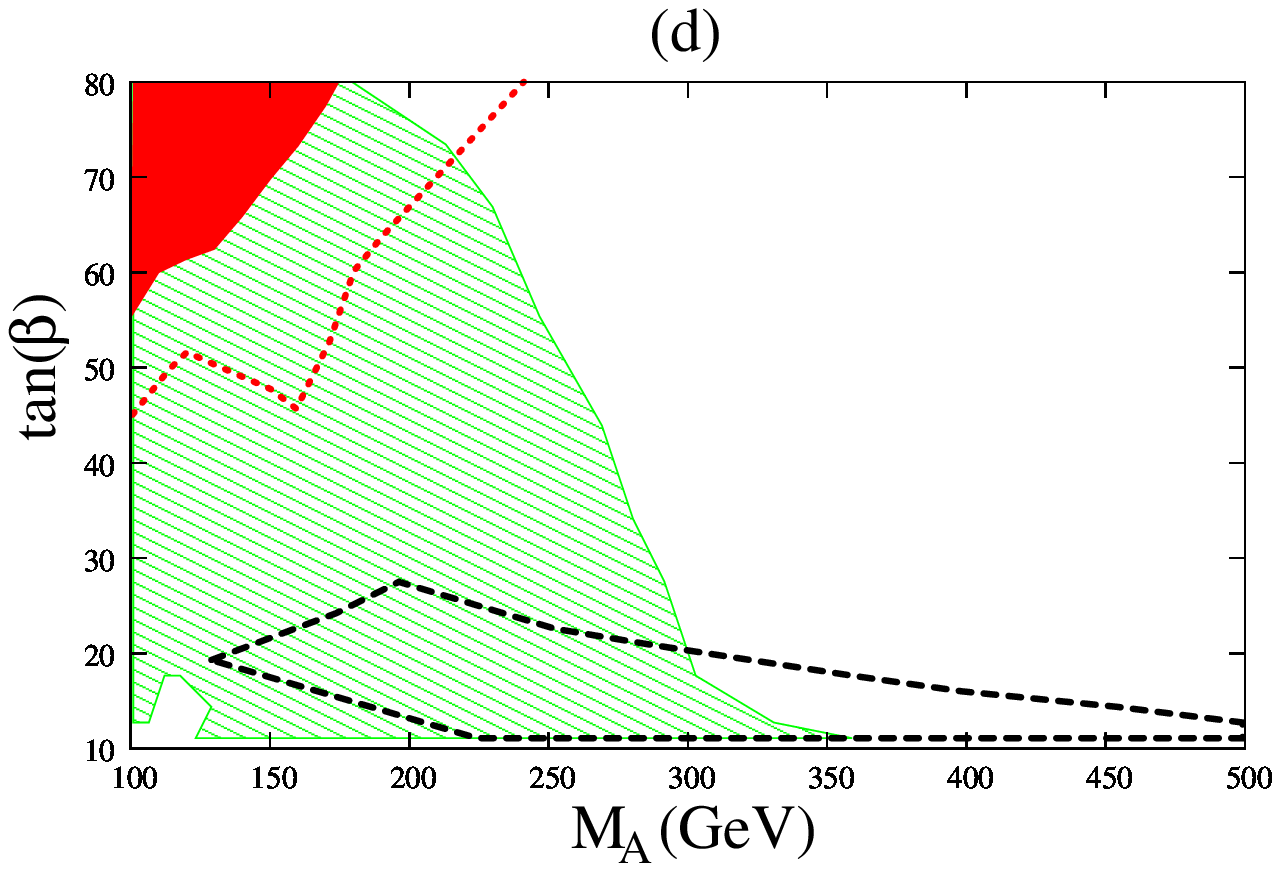}}
\end{center}
\caption{The red (grey) region, in all four figures, is excluded by the CDF 
experiment's search for non-standard Higgs bosons in the inclusive $A \to 
\tau^+ \tau^-$ channel at 1~fb$^{-1}$ luminosity.  {
The \emph{dotted line} shows the corresponding 
D0 excluded region at 1~fb$^{-1}$. (a) The \emph{solid} and \emph{dashed lines}
represent the future reach for the Tevatron (at 4 fb$^{-1}$)and LHC 
(at 10 fb$^{-1}$ for $B_s \to \mu^+ \mu^-$ and at 30 fb$^{-1}$ for 
$A \to \tau^+ \tau^-$) respectively, where  
the \emph{red (dark gray) lines} correspond to the non-standard 
Higgs search reaches in the $H\rightarrow \tau \tau$ channel while the 
\emph{black lines} are the projected $\mathcal{BR}(B_s\rightarrow \mu^+ \mu^-)$
bounds for $\mu = -100 $~GeV, $X_t = 2.4$~TeV, $M_{SUSY} = 1$~TeV and $M_3 = 
0.8$~TeV.} The green (gray) hatched regions are those allowed by the present 
B-physics 
constraints on the $B_u \to \tau \nu$ $b \rightarrow s \gamma$ and $B_s \to 
\mu^+ \mu^-$ branching ratios. (b) and (c)~For the same SUSY mass parameters
the yellow (light gray) area is the 5$\sigma$ discovery region in the
$h \rightarrow \gamma \gamma$ channel, while the green (gray) hatched area is
the same for the $h \to \tau \tau$ channel for the CMS and ATLAS experiments 
respectively at 
30~fb$^{-1}$. (d) Green (gray) hatched region is the 3$\sigma$ evidence region
for the SM-like Higgs searches (at 4~fb$^{-1}$) at the Tevatron. (b)--(d) 
The areas surrounded by the dashed black curves correspond to the regions 
allowed by present B-physics constraints.}
\label{xt24mu1n:fig}
\end{figure}

In this section we shall use the above B physics limits and Higgs search
capabilities to put constraints on the allowed regions of MSSM parameter space.
In particular we project these constraints onto the $M_A-\tan \beta$ plane.
We also assume that all the squark masses are uniform and denoted by 
$M_{SUSY}$, $2M_1=M_2=500$~GeV and we use the central value for the top-quark
measured, at the Tevatron to be 
$m_t = 170.9 \pm 1.8$~GeV~\cite{Brubaker:2006xn}. 
Within this framework we study four benchmark 
scenarios by varying the parameters $\mu$, $X_t = A_t - \mu/\tan \beta$, 
$M_{SUSY}$ and $M_3$. We numerically calculate the ratio $r$, defined for 
non-standard Higgs searches in Eq.~(\ref{rratio_nsmh:eq}), using the CPsuperH 
program~\cite{Lee:2003nt}.  {
To estimate the present excluded region and the projected Tevatron reach
we used the 1~fb$^{-1}$ CDF results presented in Ref.~\cite{cdf1fbhtautau}, 
the projected 4~fb$^{-1}$ curves from Ref.~\cite{TeVBstomumu} and the 
1~fb$^{-1}$ D0 results from 
Ref.~\cite{D01fbhtautau} for the maximal mixing scenario with 
$\mu \sim -200$~GeV.} To estimate the LHC reach we used the results for the 
maximal mixing scenario with $\mu \sim -200 $~GeV in Fig. 6 of 
Ref.~\cite{Carena:2005ek}, which is based on the study in 
Ref.~\cite{Abdullin:2005yn}.  
Using Eq.~(\ref{rratio_nsmh:eq}),
each of these curves are rescaled for each of the different parametric 
scenarios we consider in this paper. Let us stress that 
the results of Ref.~\cite{Abdullin:2005yn}, we are using, are in reasonably 
good agreement with the latest CMS studies for different $\tau$ decay final 
states, which include a
full detector simulation~\cite{cmsnote1,cmsnote2,cmsnote3,gennai:07aa}.

For the SM-like Higgs searches at 30~fb$^{-1}$, we used the CMS and the 
ATLAS studies shown in Ref.~\cite{nikitenko_cms,Abdullin:2005yn} to estimate 
the signal significance in the $h \to \tau \tau$ and $h \to \gamma \gamma$ 
channel. We used CPsuperH~\cite{Lee:2003nt} to calculate the relevant branching
ratios and couplings needed to estimate the value of $R$ in 
Eq.~(\ref{rratio:eq}). For the Tevatron searches we used the updated values of 
the luminosity needed to discover a Standard Model Higgs, from 
Ref.~{\cite{Babukhadia:2003zu}}, to estimate the variation of signal 
significance with respect to SM Higgs mass at 4~fb$^{-1}$ for each experiment.
The projections at the Tevatron assume an improvement 
in the sensitivity of detectors along with a basic increase in the 
luminosity~\cite{Babukhadia:2003zu}.

Before presenting our analysis, let us stress that,
from the form of the double penguin contribution to $\Delta M_s$ in 
Eq.~(\ref{dmsdp:eq}) and
the large $\tan \beta$ contribution to $\mathcal{BR}(B_s \to \mu^+ \mu^-)$ in 
Eq.~(\ref{BRbsmumu:eq}), it is clear that the two quantities are greatly 
correlated. As we shown in Ref.~\cite{Buras:2002wq,Buras:2002vd} for the case 
of uniform squark masses, Eq.~(\ref{BRbsmumu:eq}) and Eq.~(\ref{dmsdp:eq}) imply
that 
\begin{eqnarray}
\frac{|(\Delta M_s)_{DP}^{SUSY}|}{\mathcal{BR}(B_s \to \mu^+ \mu^-)_{SUSY}}  
\sim \frac{0.034 {\rm (ps)}^{-1}}{10^{-7}} \frac{M_A^2}{M_W^2} 
\left(\frac{50}{\tan \beta}\right)^2.\label{dmsbsbsmumubound:eq}
\end{eqnarray}
Notice that the only SUSY parameters this ratio depends on are $M_A$ and $\tan 
\beta$. 
Considering the present experimental limit on  $\mathcal{BR}(B_s \to \mu^+ 
\mu^-)$ in Eq.~(\ref{bsmumubound:eq}), we showed in Ref.~\cite{Carena:2006ai} 
that, as is apparent in Eq.~(\ref{dmsbsbsmumubound:eq}), the double penguin 
contributions to $\Delta M_s$ can be at most a few ps$^{-1}$ for $M_A < 1$~TeV.
As these corrections are negative with respect to 
the SM contribution, they make the theoretical predictions agree slightly
better with the experimentally measured value.  {
However given that the theoretical 
errors in Eq.~(\ref{deltamsbound:eq}) and Eq.~(\ref{deltamsbounda:eq}) are 
large and the SUSY contributions are small, the $\Delta M_s$ measurement only 
puts a very weak constraint on Higgs searches once the $B_s \to \mu^+
\mu^-$ bound is imposed.}

\subsection{Large to moderate $X_t$ and small $\mu$}

This scenario is a modified version of the one called maximal 
mixing because we chose the sign of $A_t M_3$ to be negative. This choice of 
sign tends to reduce the value of the SM-like Higgs mass making it easier for 
the Tevatron collider to possibly probe this scenario.  
On the other hand the change in the sign of $M_3$ with respect to that in the 
maximal mixing scenario~\cite{Carena:2005ek} does not significantly affect 
B-physics constraints and the non-standard Higgs boson search limits, as can be
seen in Fig.9(a) of Ref.~\cite{Carena:2006ai}.  {The SM-like 
Higgs mass depends strongly on the stop mixing parameter $X_t$, and it attains 
its maximum value for $X_t \sim \sqrt{6} M_{SUSY} = 2.4$~TeV.}
For these values of $X_t$, small $\mu$ and small 
$M_A$, which can be probed at the Tevatron, we need the sign of $\mu A_t$ to be
negative so that the stop-chargino contribution to $b \to s \gamma$ amplitude 
in Eq.~(\ref{bsgchi:eq}) cancels against that of the charged Higgs in 
Eq.~(\ref{bsgh+:eq})~\cite{Carena:2006ai}. The $B_s \to \mu^+ \mu^-$ constraint
in this scenario is quite strong because the $B_s \to \mu^+ \mu^-$ branching
ratio in Eq.~(\ref{BRbsmumu:eq}) is proportional to $A_t$, which is large, and
in the denominator the factor $1+\epsilon_3 \tan \beta \sim 1$, as the 
$\epsilon_3$ loop-factor is small. The 
$B_u \to \tau \nu$ constraint has two allowed regions related to the two
possible signs of the amplitude, as can be seen in Eq.~(\ref{rbtaunu:eq}). 
At low values of $\tan \beta$ and large values of $M_A$ the SM 
contribution dominates, while at complementary values of $M_A$ and $\tan \beta$
the SUSY contribution dominates. 

In Fig.~\ref{xt24mu1n:fig}~(a) the present limit on the 
$B_s \rightarrow \mu^+ \mu^-$, and the measurements of the $b \to s \gamma$ and
$B_u \to \tau \nu$ decay rates allow the green (gray) hatched region for 
$X_t=2.4$~TeV, $M_3 = -800$~GeV, $M_{SUSY}=1$~TeV and $\mu=-100$~GeV.
 {The red (dark gray) region is excluded by the CDF experiment's 
non-standard Higgs search in the inclusive $\tau^+ \tau^-$ decay mode.} The 
dotted red (dark grey) is
the corresponding excluded region according to the D0 collaboration.
The red (dark gray) solid and dashed curves show the regions that can be 
excluded by non-standard Higgs searches at
the Tevatron for a future luminosity of 4~fb$^{-1}$ and at the LHC for a  
luminosity of 30~fb$^{-1}$ respectively.
The black solid and dashed curves corresponds to the future $B_s 
\rightarrow \mu^+ \mu^-$ limits for the Tevatron at a luminosity of 4~fb$^{-1}$
and the LHC at a luminosity of 10~fb$^{-1}$ shown in 
Eq.~(\ref{bsmumuboundf:eq}) and Eq.~(\ref{bsmumuboundLHC:eq}) respectively.
 {A reach similar to Eq.(\ref{bsmumuboundLHC:eq}) and comparable 
to the standard model prediction is expected at LHCb with only a few fb$^{-1}$ 
of data~\cite{LHCB}.} 
As the B-physics allowed region corresponds to large values of $M_A$ and small 
values of $\tan \beta$, the SM contribution to the amplitude of the 
$B_u \to \tau \nu$ process is larger than the SUSY contribution to the same 
amplitude. The region where the SUSY contribution to the amplitude of the 
$B_u \to \tau \nu$ process is larger than the SM contribution is excluded by 
the present bounds on the $B_s \to \mu^+ \mu^-$ branching ratio in 
Eq.~(\ref{bsmumubound:eq}).

As we found in Ref.~\cite{Carena:2006ai} the maximal mixing scenario is 
strongly constrained by B-physics and the addition of the $B_u \to \tau \nu$
limit makes these constraints even stronger.
For these values of SUSY parameters 
B-physics constraints prefer low to moderate values of $\tan \beta$. In 
addition the Tevatron will find it difficult to discover a non-standard Higgs
boson for this scenario. Moreover, the LHC at a luminosity of 30~fb$^{-1}$ will
only be able to probe a very small portion of the B-physics allowed parameter 
space in the $A/H \to \tau \tau$ channel.

\begin{figure}
\begin{center}
\resizebox{7.cm}{!}{\includegraphics{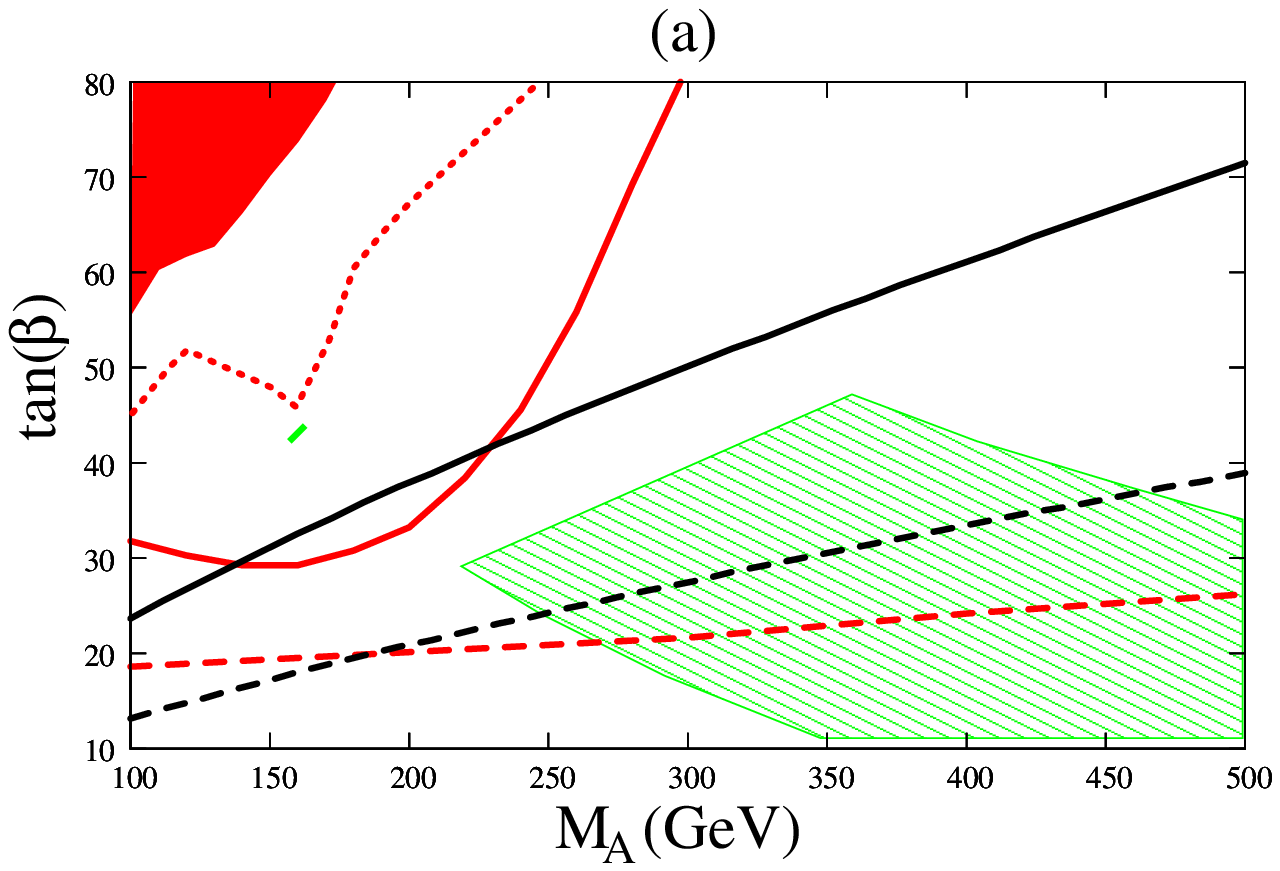}}
\resizebox{7.cm}{!}{\includegraphics{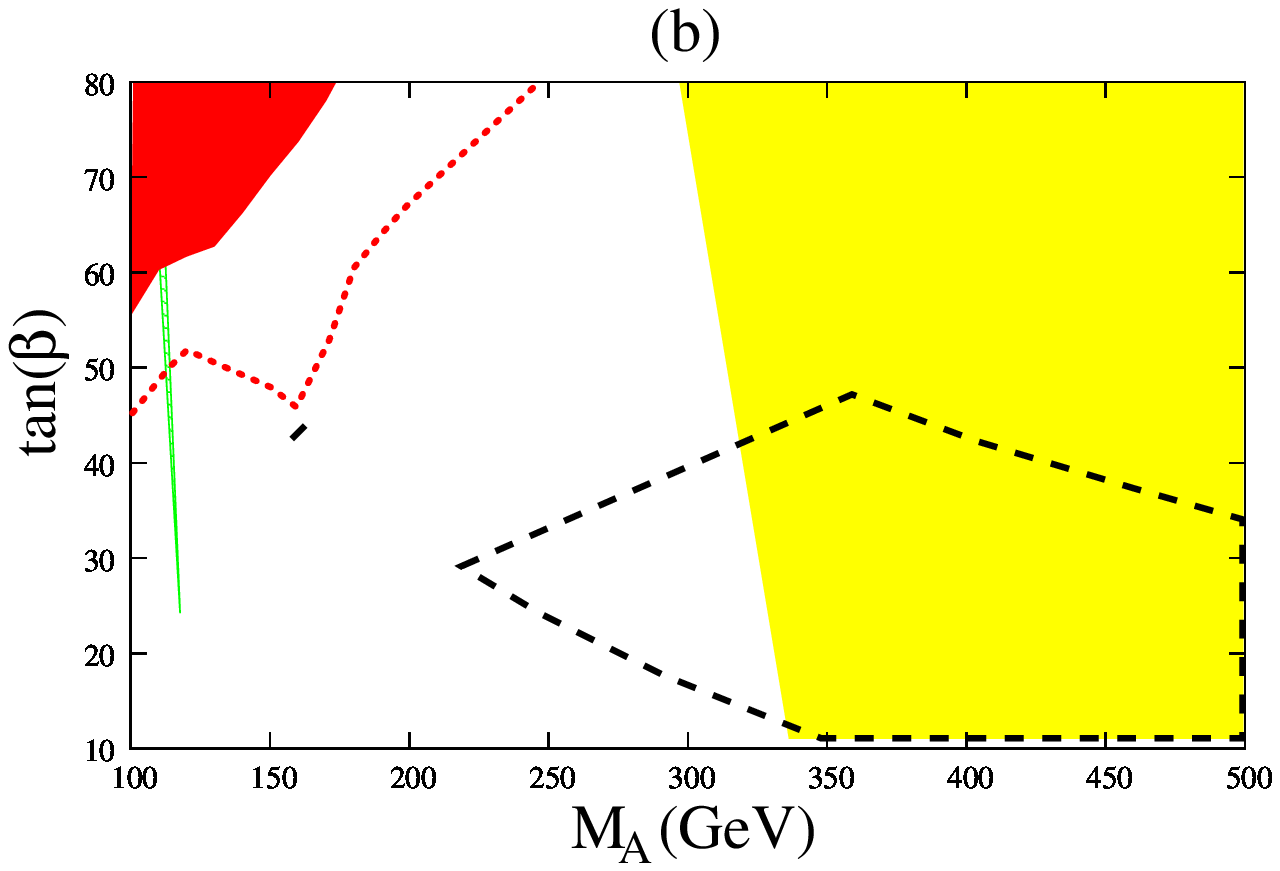}}
\resizebox{7.cm}{!}{\includegraphics{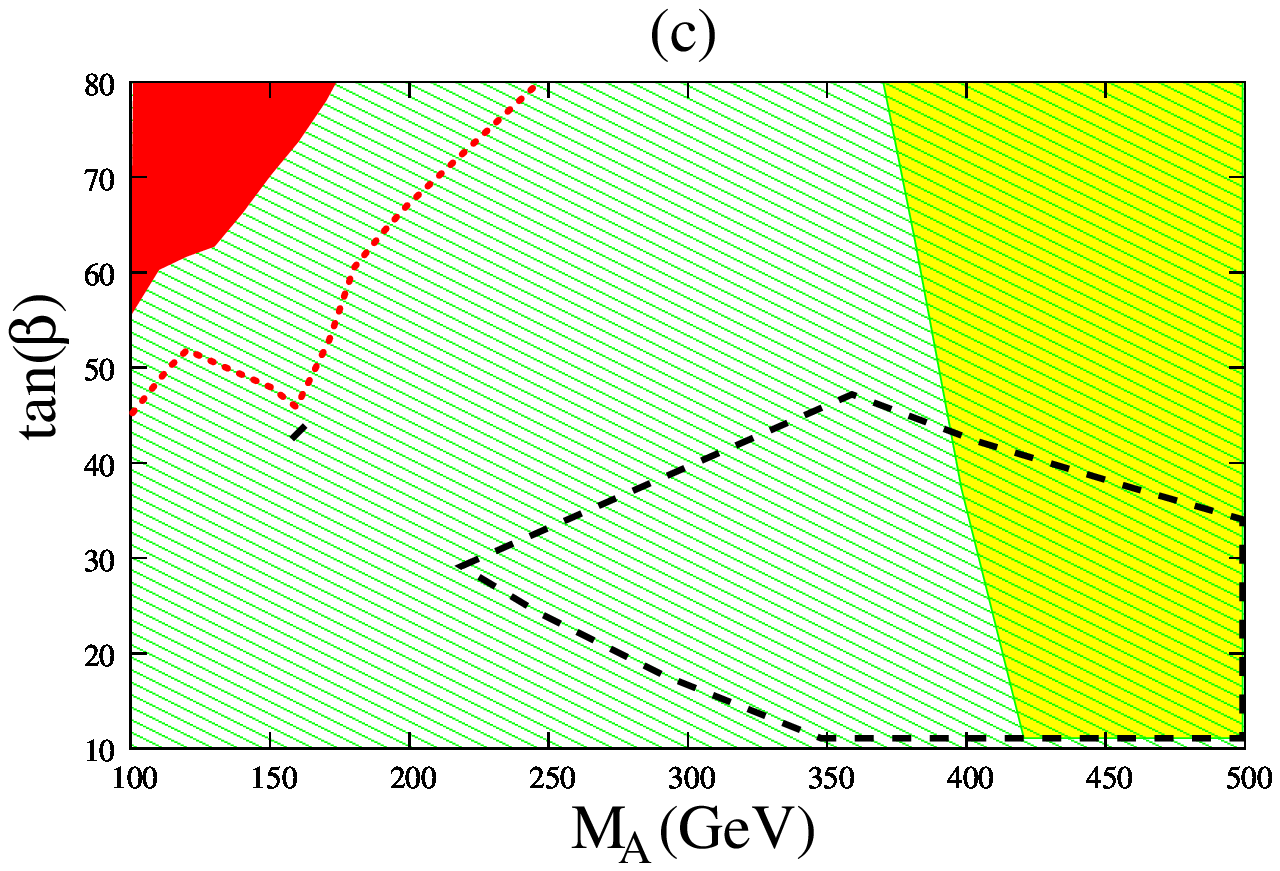}}
\resizebox{7.cm}{!}{\includegraphics{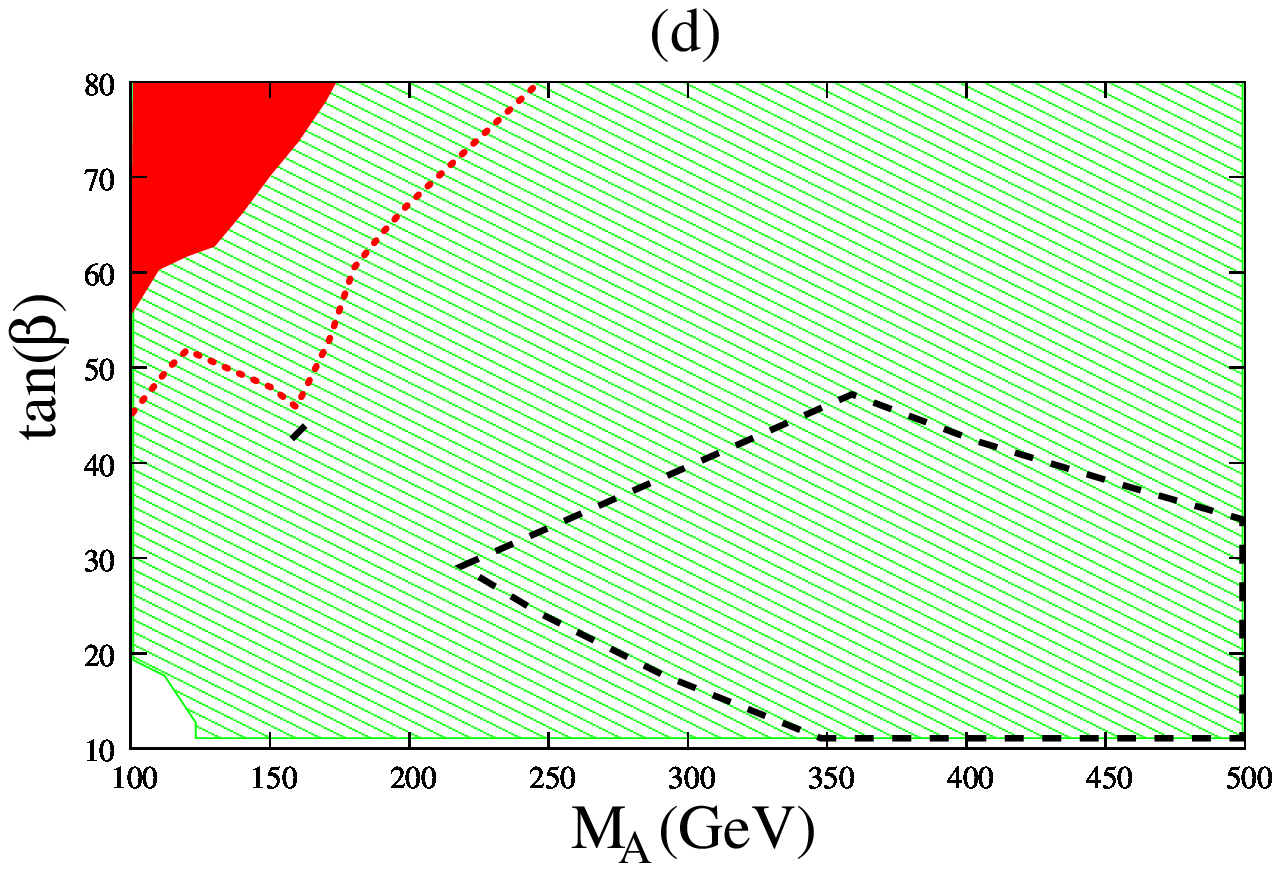}}
\end{center}
\caption{(a)--(d) The lines and the colors correspond to the same quantities
as in Fig.~(\ref{xt24mu1n:fig}), where the SUSY parameters are the same 
except for $X_t = 1$~TeV.}
\label{xt10mu1n:fig}
\end{figure}

In Fig.~\ref{xt24mu1n:fig}~(b and c) we show the parts of the $M_A \-- \tan 
\beta$ that can be probed in Standard Model Higgs searches at the CMS and 
ATLAS experiments, respectively.
The yellow (light gray) regions are those that can be probed in 
$h \to \gamma \gamma$ channel while the green (dark gray) hatched regions can 
be probed in $h \to \tau \tau$ channel with a luminosity of 30~fb$^{-1}$ at 
5~$\sigma$. Present available studies with the ATLAS detector show that it 
will be able to probe all of the B-physics 
allowed region. According to the new analysis shown in 
Ref.~\cite{nikitenko_cms}, the CMS detector may not be able to probe the region
of moderate $M_A$ in the $h \to \tau \tau$ channel.
However due to a significant improvement in the CMS sensitivity in the
$\gamma \gamma$ channel a large portion of the B-physics allowed region can 
still be probed. If the sign of $A_t M_3$ were positive the qualitative 
features of the CMS reach and ATLAS reach would remain the same.

In Fig.~\ref{xt24mu1n:fig}~(d) we show the region of the $M_A \-- \tan \beta$ 
plane 
that the Tevatron can probe in the $h \to b \bar{b}$ channel with a luminosity 
of 4~fb$^{-1}$ per experiment and a signal significance of 3 standard 
deviations. 
For the modified maximal mixing scenario the region that can be probed is 
relatively large compared to the standard one~\cite{Carena:2005ek,Aglietti:2006ne}, because
the sign of $A_t M_3$ is negative. For negative $A_t M_3$ the maximum SM-like 
Higgs boson mass is approximately $\sim 125$~GeV compared to the standard 
maximal mixing scenario which has $130$~GeV as the maximum Higgs 
mass~\cite{Lee:2003nt}.

In Fig.~\ref{xt10mu1n:fig} we show the effect of going to a lower value of 
stop mixing parameter $X_t = 1$~TeV. There are two disconnected B-physics 
allowed regions for these
SUSY parameters shown in Fig.~\ref{xt10mu1n:fig}~(a). There is a tiny upper 
region at around $(M_A, \tan \beta) \sim (150 \mbox{ GeV}, 43)$ and a much 
larger lower $\tan \beta$ region where all the B physics constraints are just 
satisfied. In the upper region the SUSY 
contribution to the amplitude of the $B_u \to \tau \nu$ rate is larger than the
SM contribution 
to the same process, while in the lower region the opposite is true. The area
between these two regions is excluded because the ratio $R_{B\tau \nu}$ in 
Eq.~(\ref{btaunubound:eq}) is below the 2$\sigma$ bound. The reach via SM-like 
Higgs searches for these SUSY parameters, are similar to the maximal mixing 
scenario.
CMS has difficulties seeing the SM-like Higgs in part of the regions allowed by
B-physics 
constraints, but the ATLAS experiment will cover all of $M_A \-- \tan \beta$
plane. The Tevatron experiments may now cover the whole allowed region of the $M_A \-- \tan \beta$ plane at 3$\sigma$.

\subsection{Large $\mu$ and small or negligible $X_t$ }

For the minimal mixing scenario,  $X_t$ is equal to zero and the 
chargino-stop contribution to the $b \to s \gamma$ process is small.
Due to a reasonable agreement between the Standard Model prediction and the 
experimental measurement of the $b \to s \gamma$ rate, we need the charged 
Higgs contribution in Eq.~(\ref{bsgh+:eq}) to be small. For a light charged 
Higgs, this requirement
can be achieved by going to large values of $\mu$, $M_3$ and $\tan \beta$
because of a cancellation between the tree-level term and the loop induced
term in Eq.~(\ref{bsgh+:eq}). Since $A_t$ is small, the $B_s \to \mu^+ \mu^-$  
limit puts a weak constraint on the $M_A \-- \tan \beta$ plane. 
Additionally, for these values of parameters the usual bound on $\tan \beta$ 
that comes from requiring that $y_b$ be perturbative up to the GUT scale may be
relaxed: Since the bottom Yukawa has the form 
\begin{eqnarray}
y_b \simeq \frac{\sqrt{2} m_b \tan\beta}{v (1+\epsilon_3 \tan \beta)}
\end{eqnarray}
and as $\epsilon_3 \tan\beta$ needs to be real, positive and of order one,
for the above cancellation in the charged Higgs amplitude to occur\footnote{
An exact cancellation is not needed due to the theoretical and experimental 
uncertainties so a small phase is also allowed}, the denominator suppresses
the bottom Yukawa coupling for large values of $\tan \beta$.

The SM-like Higgs searches put an interesting constraint on scenarios with 
large values of $|\mu|$ and small values of $X_t$, since unless $M_{SUSY}$ is 
sufficiently large the SM-like Higgs 
mass tends to be below the LEP bound of $114.4$~GeV. 
The impact of the LEP bound on the excluded region in the $M_A \-- \tan \beta$ 
plane is very sensitive to 
$\mu$, $M_{SUSY}$ and the top mass.
For instance, for
$M_{SUSY} \sim 1$~TeV this scenario is highly constrained by the LEP bounds on
the SM-like Higgs mass, but increasing $M_{SUSY}$ to $2$~TeV is sufficient to 
avoid this constraint~\cite{Carena:2002qg}.

\begin{figure}[ht]
\begin{center}
\resizebox{7.cm}{!}{\includegraphics{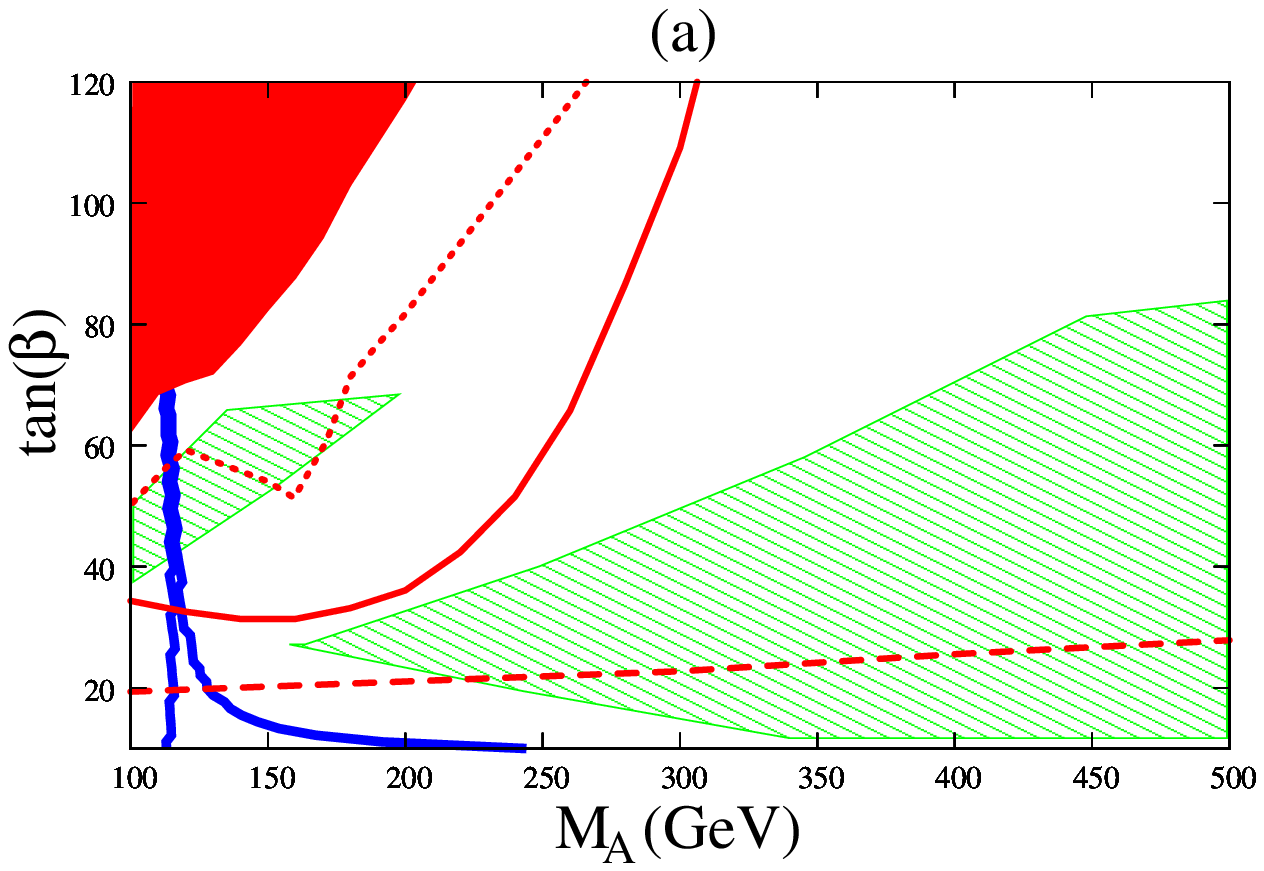}}
\resizebox{7.cm}{!}{\includegraphics{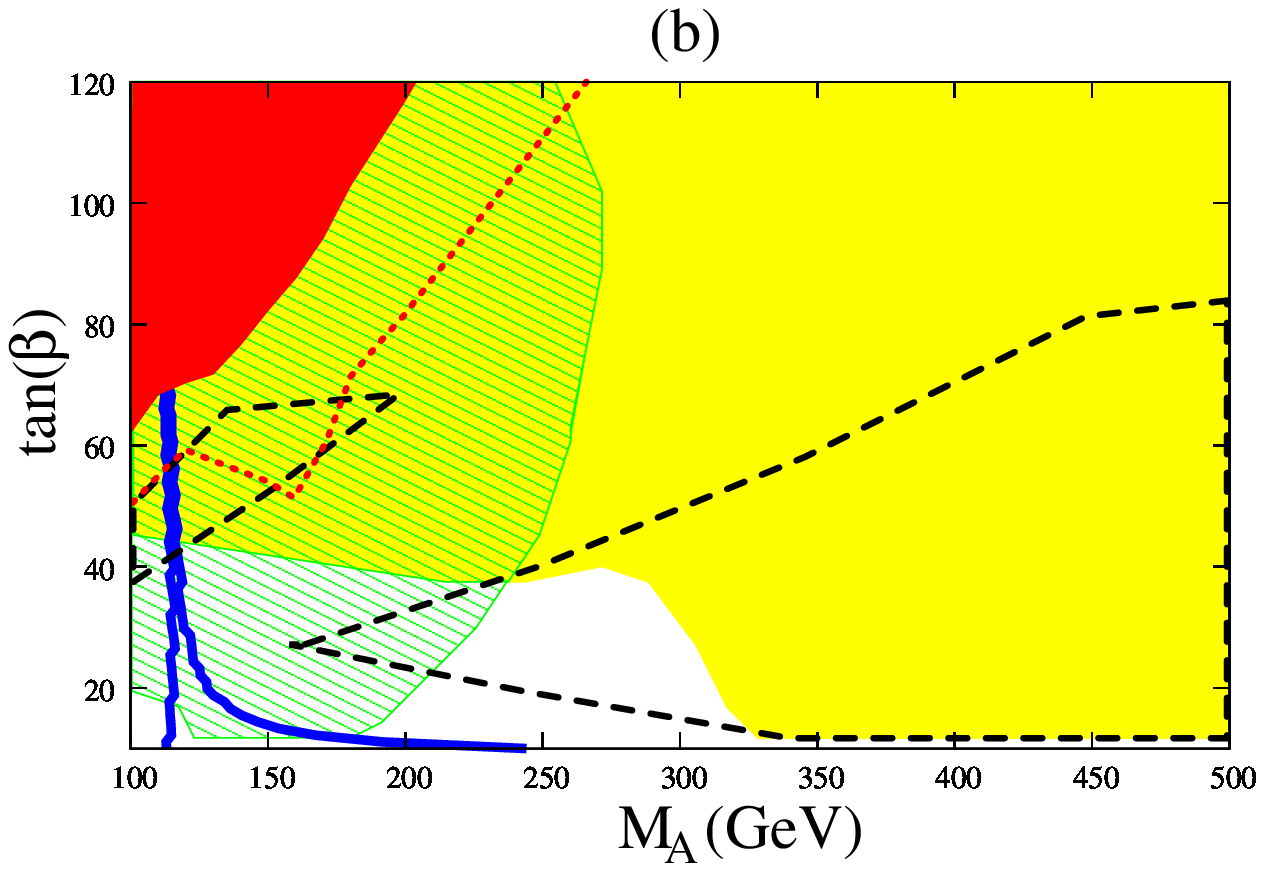}}
\resizebox{7.cm}{!}{\includegraphics{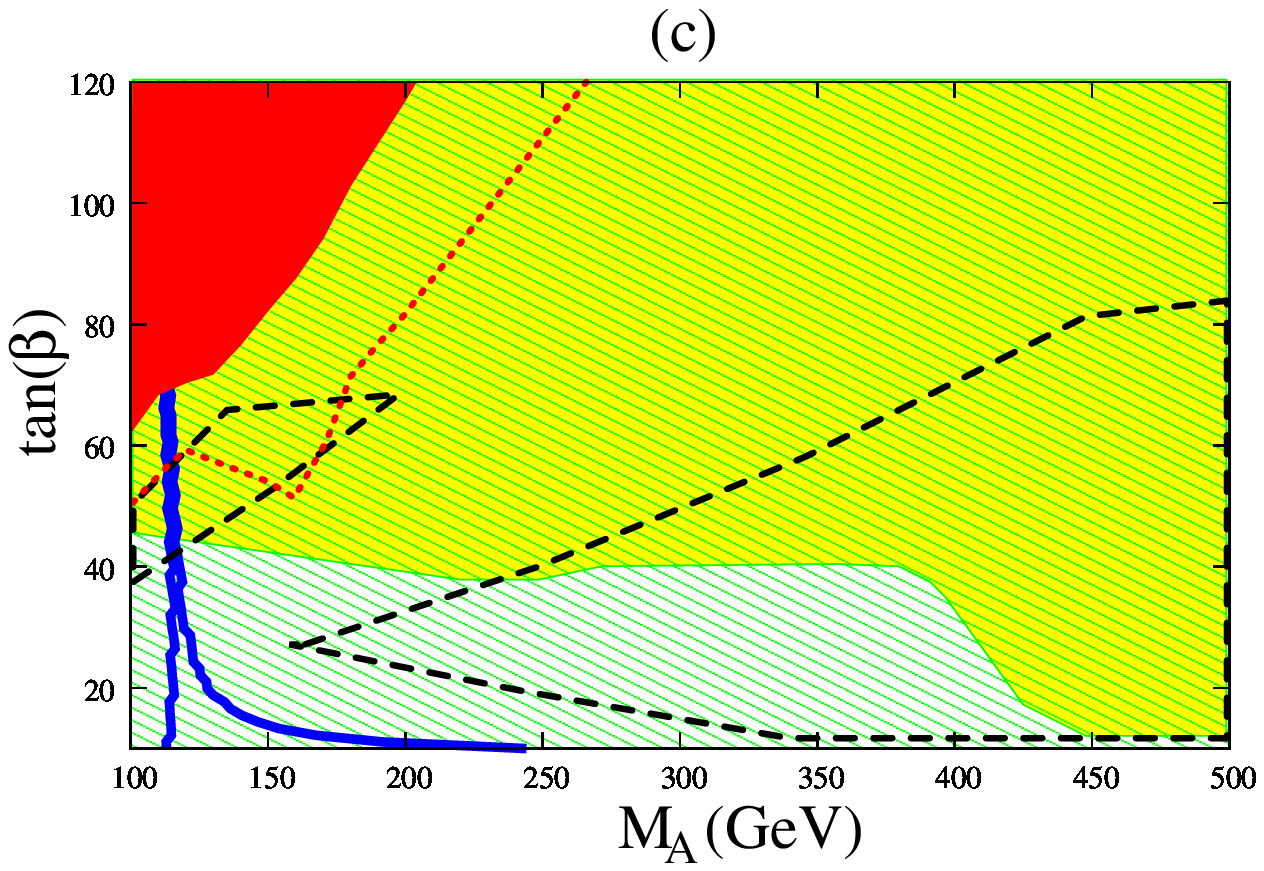}}
\resizebox{7.cm}{!}{\includegraphics{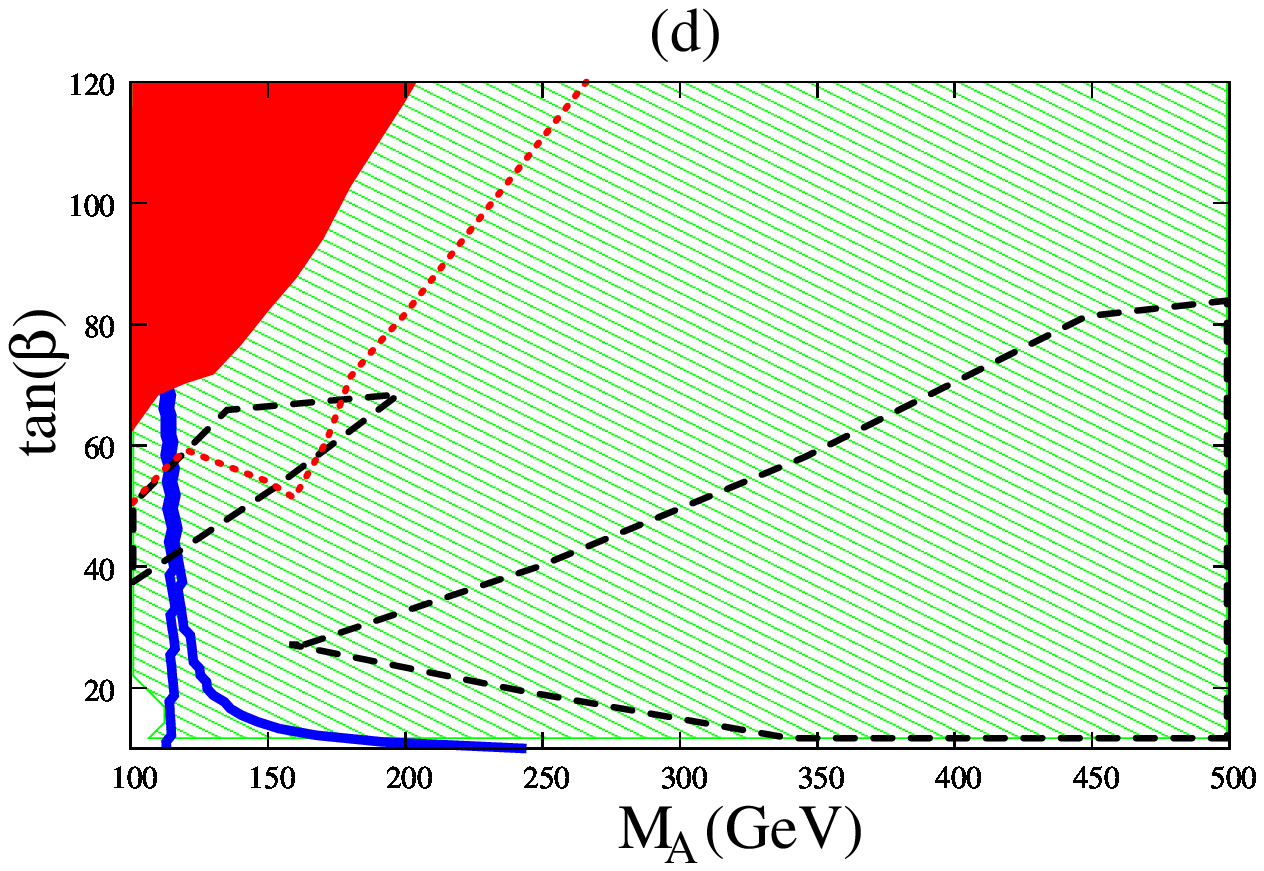}}
\end{center}
\caption{(a)--(d) The lines and the colors correspond to the same quantities
as in Fig.~(\ref{xt24mu1n:fig}), where the SUSY parameters are the same 
except for $X_t = 0$~GeV, $\mu = 1.5 \; M_{SUSY}$ and $M_{SUSY} = 2$~TeV. 
The  region below the blue (black) solid
line corresponds to the area excluded by the LEP bound on the 
SM-like Higgs boson for $m_t = 170.9$~GeV.}
\label{xt0mu15:fig}
\end{figure}

 The corresponding results for $M_{SUSY} = 2$~TeV are shown in 
Fig.~\ref{xt0mu15:fig}. We have previously analyzed  
this scenario in Ref.~\cite{Carena:2006ai} without adding the $B_u \to \tau 
\nu$ constraints. In Fig.~\ref{xt0mu15:fig} we see that the addition of this
new constraint excludes the diagonal region with corners $(100 \mbox{ GeV}, 38
), (155 \mbox{ GeV}, 28),(450 \mbox{ GeV}, 80)$ and $(190 \mbox{ GeV}, 65)$
for the parameters $\mu = 1.5 \; M_{SUSY}$ and $M_3=0.8 \; M_{SUSY}$.
In 
Fig.~\ref{xt0mu15:fig}~(a) we show the effect of the LEP bound on the B-physics
allowed regions.
The region below the blue (black) solid line shows the area excluded by the 
LEP bound in the $M_A \-- \tan \beta$ plane.

From Fig.~\ref{xt0mu15:fig}~(b) and (c) it is clear that the 
CMS and ATLAS experiment can probe most of the allowed B-physics regions of the $M_A \-- 
\tan \beta$ plane, using SM-like Higgs searches in the $h \to \gamma \gamma$ 
and the $h \to \tau \tau$ channels. CMS has an inaccessible 
region at large $M_A$ in the $\tau \tau$-channel because in this region the
$\tau$ Yukawa coupling is only slightly above the standard model value and 
according to 
Ref.~\cite{nikitenko_cms} CMS does not have a 5$\sigma$ signal
significance with 30~fb$^{-1}$ of data for any standard model Higgs mass.
However, given that the Higgs mass and the
$h \to \tau \tau$ coupling vary smoothly with $M_A$ and $\tan 
\beta$ the discovery potential is also above 4$\sigma$ 
for most of the region that appears inaccessible in Fig.~\ref{xt0mu15:fig}~(b).
Again, at 4~fb$^{-1}$ the Tevatron could
have a 3$\sigma$ evidence over most of the parameter space allowed by 
B-physics and the LEP Higgs mass bound.

We would like to stress that the B physics and the LEP excluded 
regions, for the minimal mixing scenario, allow a clear region of $M_A = 130 
\--  170$~GeV and $\tan \beta = 50 \-- 70$. These values are easily within the
Tevatron's sensitivity region for non-standard Higgs searches in the $\tau \tau
$ channel. In addition, the SM-like Higgs boson mass is close to the current 
limit and therefore should be visible at the Tevatron at the 3~$\sigma$ level 
with an increase in sensitivity and luminosity. Both CDF and D0 collaborations 
have recently
made public their findings in the inclusive $A \to \tau \tau$ channel at a 
luminosity of  1~fb~$^{-1}$. The CDF experiment finds a 
slight excess~\cite{cdf1fbhtautau} while the D0 experiment~\cite{D01fbhtautau} 
finds a reduction in the signal for the same values of the the $\tau 
\tau$ visible mass. The D0 limit further limits the upper 
B-physics allowed region to values of $M_A = 130 \-- 150$~GeV and $\tan 
\beta \sim 55$. 

This scenario can be relatively insensitive to small changes in the value
of $X_t$. It would seem that increasing the value of $X_t$ would make
the $B_s \to \mu^+ \mu^-$ constraint extremely strong. However, 
there is a $1/\mu^2$ dependence from the $(1+\epsilon_0^3)(1+\epsilon_3)$ 
factor in the denominator of Eq.~(\ref{BRbsmumu:eq}) and only a linear $\mu$ 
dependence in its numerator. Thus as long as the loop factors $\epsilon$ 
are positive and $\mu$ is large, even moderate values of $X_t$ do not 
strengthen the $B_s \to \mu^+ \mu^-$ constraint. Additionally at large values
of $\mu$, $M_3$ and $\tan \beta$ the charged Higgs contribution to the $b \to s 
\gamma$ amplitude in Eq.~(\ref{bsgh+:eq}) may have the opposite sign to the SM 
one, a novel result that only occurs for this range of parameters. In this 
region of parameter space, to cancel this negative charged Higgs amplitude we 
need the chargino-stop contribution in Eq.~(\ref{bsgchi:eq}) to be positive or 
the sign of $\mu A_t$ to be positive.

\subsection{Small $\alpha_{eff}$} \label{smallalpeff:sec}

This scenario was studied in Ref.~\cite{Carena:2002qg} in which the 
off-diagonal components of the CP-even Higgs mass matrix are approximately 
zero. This approximate cancellation can be achieved by making, for instance, 
the following choice of parameters
\begin{eqnarray}
\mu = 2.5 \mbox{ TeV}, \;\;\;\;\; X_t = -1200.0 \mbox{ TeV}, \;\;\;\;\; 
M_{SUSY} = 800 \mbox{GeV}, \;\;\;\;\; M_3 = 500 \mbox{GeV}.
\end{eqnarray}
 {A consequence of this cancellation is that the couplings of the
SM-like Higgs boson to the b-quarks and $\tau$-leptons are suppressed.}

\begin{figure}
\begin{center}
\resizebox{7.cm}{!}{\includegraphics{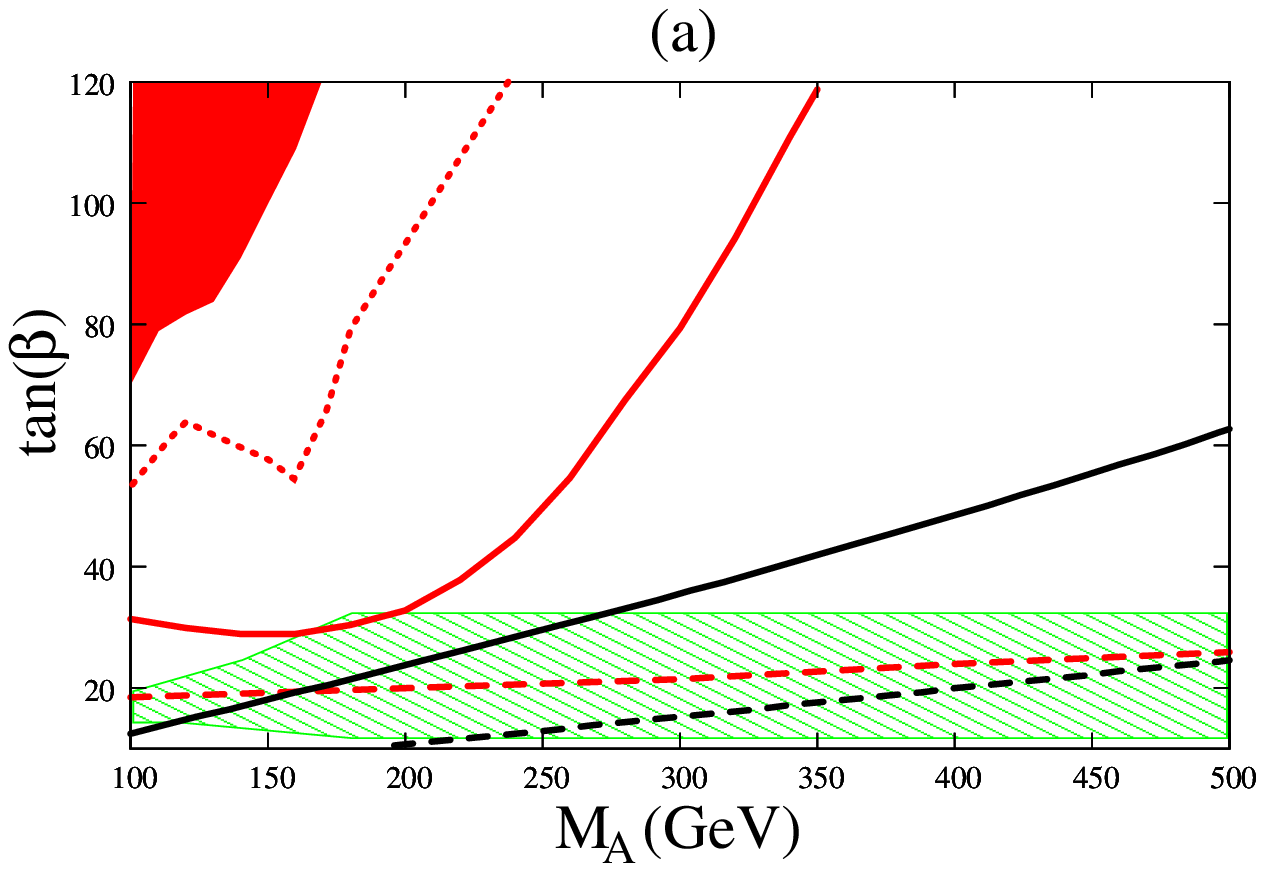}}
\resizebox{7.cm}{!}{\includegraphics{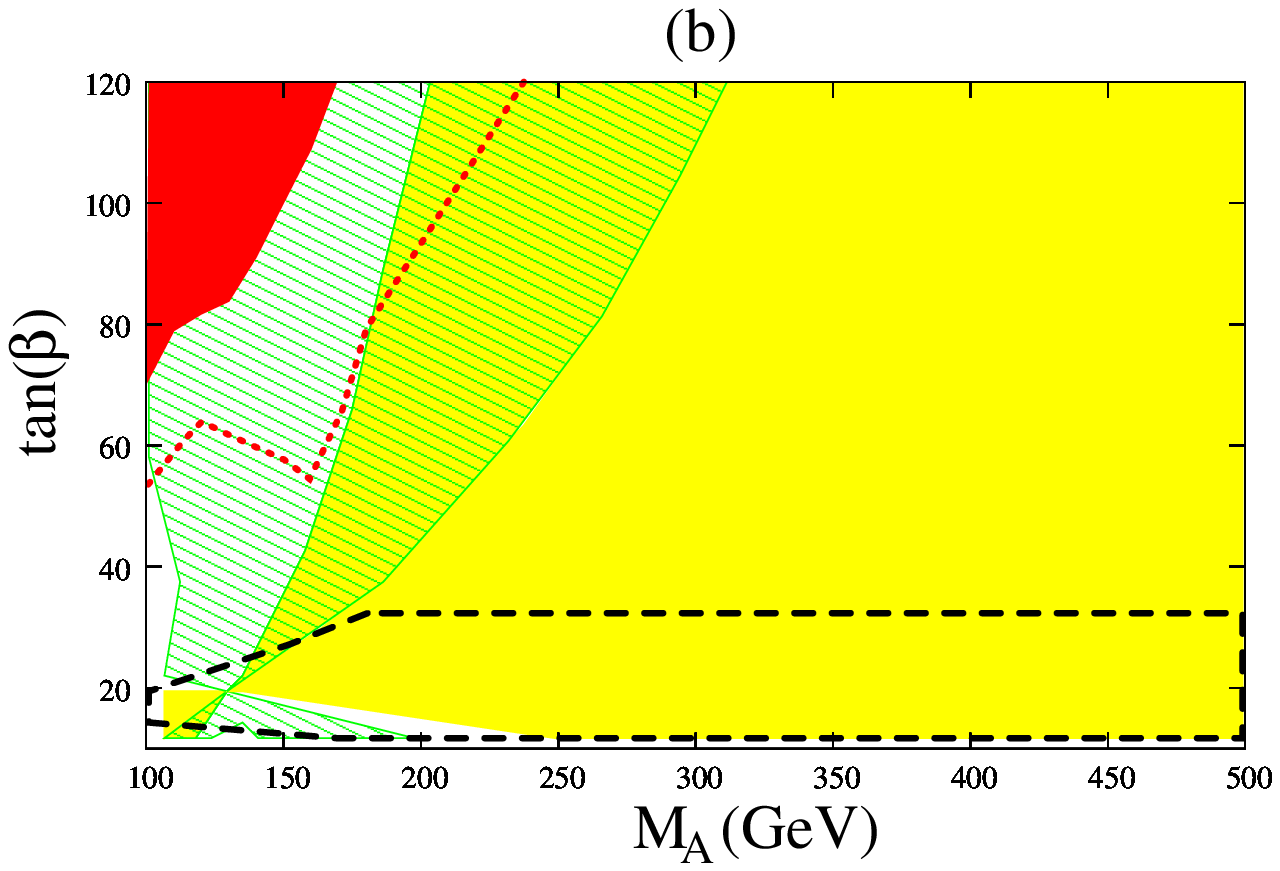}}
\resizebox{7.cm}{!}{\includegraphics{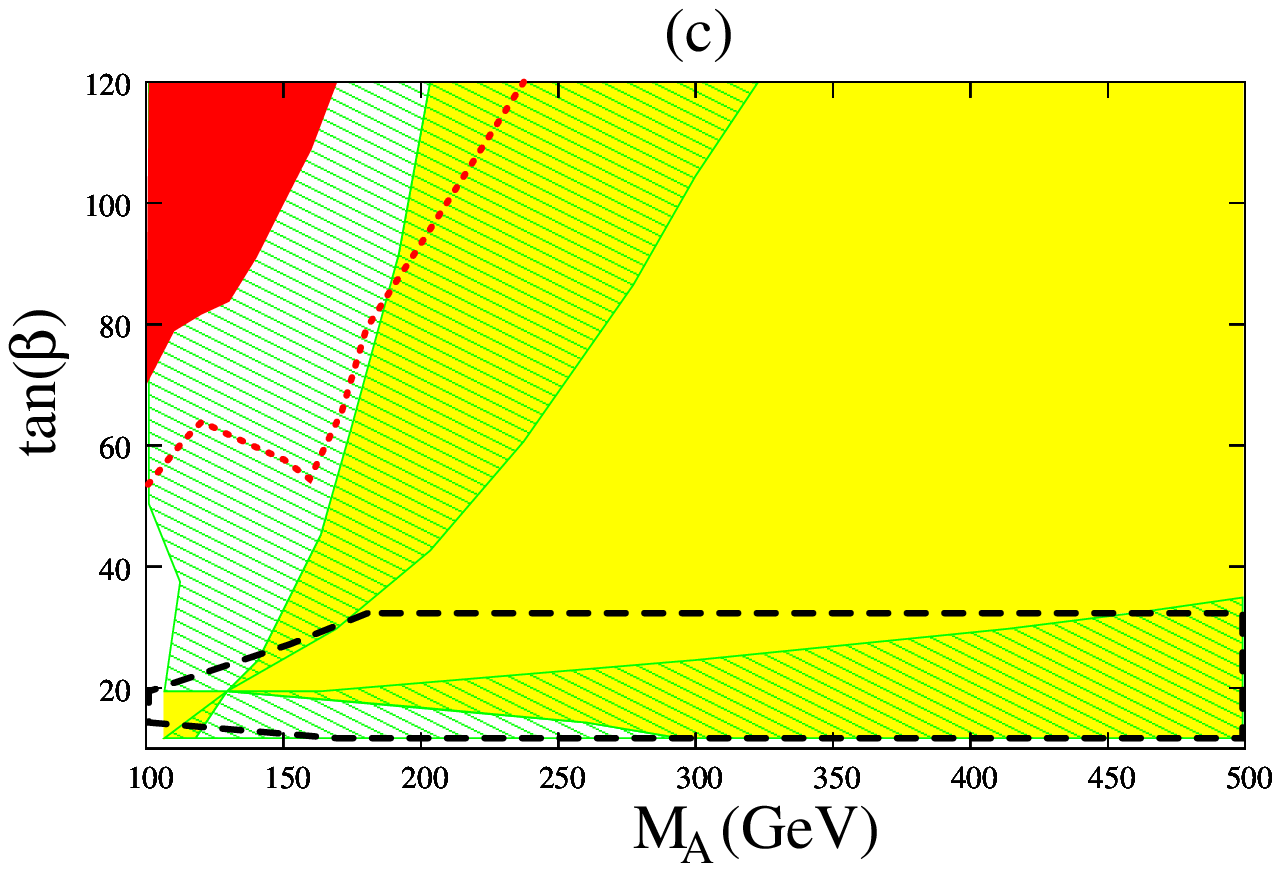}}
\resizebox{7.cm}{!}{\includegraphics{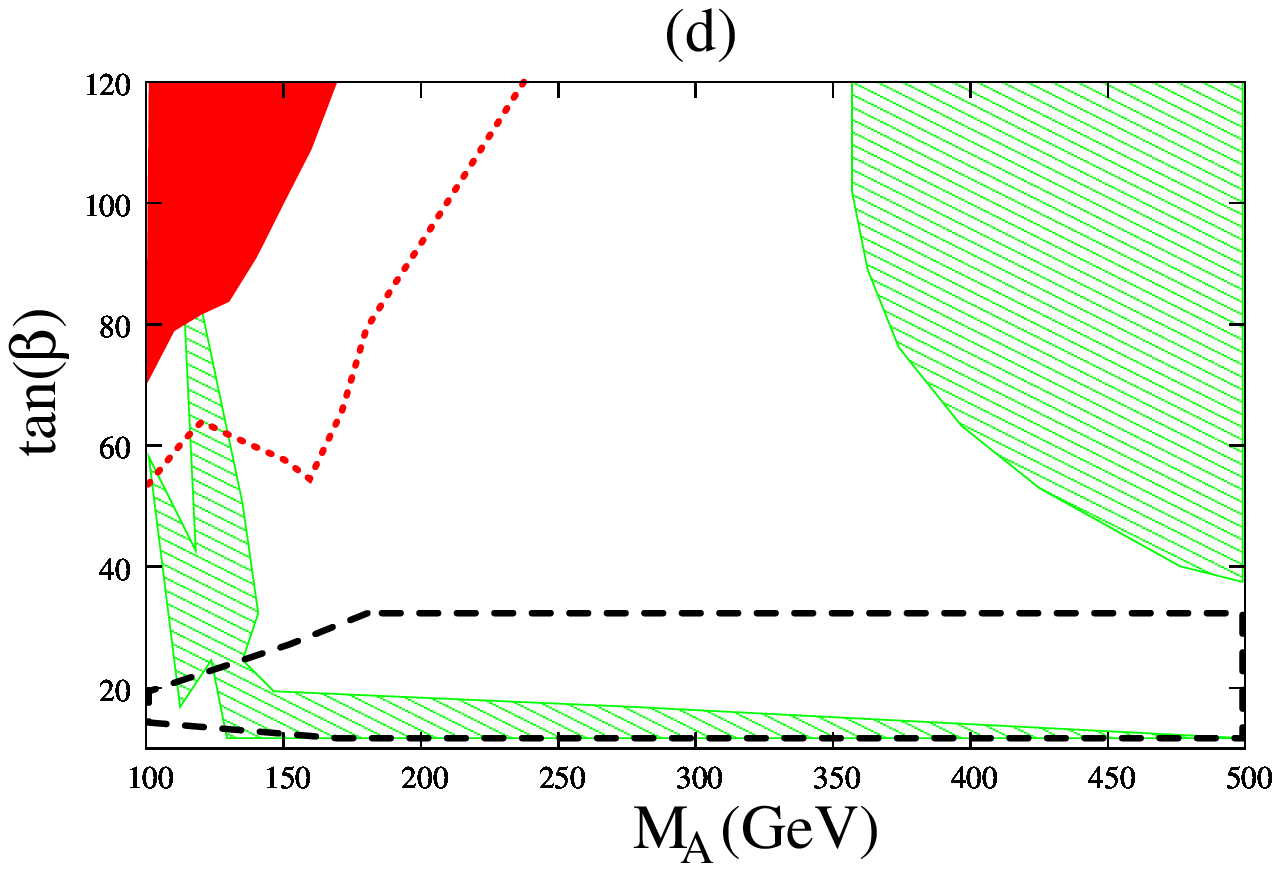}}
\end{center}
\caption{(a)--(d) The lines and the colors correspond to the same quantities
as in Fig.~(\ref{xt24mu1n:fig}), where the SUSY parameters are the same 
except for $M_3 = 500$~GeV, $M_{SUSY} = 800$~GeV, $X_t = -1.2$~TeV 
and $\mu = 2.5$~TeV.}
\label{xt12mu25:fig}
\end{figure}

 {
In Fig.~\ref{xt12mu25:fig} we present the effect of this choice of parameters
on the B-physics allowed region and on Higgs searches at the LHC and Tevatron.}
The B-physics constraints are quite severe and similar to the large $X_t$ 
scenario we discussed above. 
The $h \to \gamma \gamma$ channel for SM-like Higgs searches is enhanced 
because the $h \to \tau \bar{\tau}$ and 
$h \to b \bar{b}$ branching ratios are suppressed, leading to an enhancement
of the $h \to \gamma \gamma$ branching ratio. Therefore the CMS and 
ALTAS experiments will be able to probe a large part of the $M_A \-- \tan\beta$
plane in the $h \to \gamma \gamma$ channel. The Tevatron will not be able 
to probe most of the B-physics allowed region because of the suppression of
the $h \to b \bar{b}$ branching ratio.

\section{Conclusions}

In this article we have studied the inter-play between B-physics constraints
and Higgs searches at hadron colliders in the framework of minimal flavor 
violating SUSY models. The results we present here depend on the projected
sensitivities of the CMS and ATLAS experiments and the Tevatron collider
in the different SM-like and non-standard Higgs boson channels. The 
Tevatron projections assumed in this work~\cite{Babukhadia:2003zu}
need to be further solidified by improvements in the analyses that CDF and
D0 are performing.  Both CMS and ATLAS have recently performed 
improvements in their projections in the $\gamma \gamma$ inclusive channel and
CMS has also recently updated their $h \to \tau \tau$ vector boson fusion 
study~\cite{nikitenko_cms}. We have illustrated this interplay between
Higgs searches at hadron colliders and B-physics constraints using four 
benchmark senarios.

In particular the B-physics constraints are extremely severe for SUSY 
parameters which have large values of $X_t$ and small values of $\mu$. For 
SM-like Higgs boson searches the LHC experiments should be able to probe all 
of the allowed region of parameter space with 30~fb$^{-1}$, but
the Tevatron 
collider will have difficulties doing this with 4~fb$^{-1}$ of data. 
Discovering a SM-like Higgs boson at the CMS 
experiment with 30~fb$^{-1}$ of data will be challenging in this scenario, 
since CMS has a better sensitivity in the $h \to \gamma \gamma$ rather than in 
the $h \to \tau \tau$ channel and as the $h b \bar{b}$ and the $h \tau 
\bar{\tau}$
couplings are somewhat enhanced for moderate or small $M_A$, the $h \to \gamma 
\gamma$ branching ratio is smaller than in the SM.
On the other hand, the ATLAS experiment will easily probe the allowed region of
parameter space because the $h \to \tau \tau$ branching ratio is enhanced
for these values of SUSY parameters. The Tevatron will find it very difficult 
to detect a SM-like Higgs in this scenario because the SM-like Higgs is heavy 
and the signal significance, in the $h \to b \bar{b}$ channel, drops sharply 
with increasing Higgs mass. Additionally, in this scenario the B-physics 
constraints favor regions which have large values of $M_A$ and low values of 
$\tan \beta$ while the non-standard Higgs boson searches at hadron
colliders are less efficient in these regions.  {Therefore at a 
luminosity of 30~fb$^{-1}$ the LHC will be able to observe the SM-like Higgs, 
but may find it difficult to discover non-standard Higgs bosons.}

The B-physics constraints are far weaker for large values of $\mu$ and small
values of $X_t$ due to a suppression of SUSY contributions to
the $B_s \to \mu^+ \mu-$ and the $b \to s \gamma$ rates. At the same time
the present LEP bounds on the SM-like Higgs mass put strong constraints on
the allowed regions of parameter space, in particular for $M_{SUSY} \leq 1
$~Tev.
For the minimal mixing scenario with $M_{SUSY}=2$~TeV we have
studied, the LHC will be able to probe most of the 
B-physics allowed region in non-standard Higgs searches, for values of 
$M_A < 500$~GeV. For SM-like Higgs searches, with 30~fb$^{-1}$ of data, the CMS
collaboration should be able to probe most of the allowed regions, while the
ATLAS collaboration will be able to probe all of them. In
addition, this scenario is the most promising for the Tevatron to detect both
the SM-like Higgs and the non-standard Higgs bosons in the near future.

The final benchmark scenario we studied was that of small $\alpha_{eff}$. 
Due to 
the suppression of SM-like Higgs couplings to b-quarks and $\tau$'s, the 
$\gamma \gamma$ channel is enhanced. Due to this enhancement both the LHC 
experiments
will be able to discover the SM-like Higgs over most of the B-physics allowed 
parameter space. The Tevatron will find it difficult to detect a SM-like Higgs 
due its mass and suppressed couplings to $b \bar{b}$.

In conclusion, scenarios with lower values of stop mixing parameter $X_t$ and 
larger values of higgsino mass parameter $\mu$ 
will be easier to probe at hadron colliders through direct higgs searches of 
both standard and  non-standard Higgs bosons.  
At larger values of $X_t$, 
direct non-standard Higgs boson searches are strongly constrained by 
present bounds on B-physics observables. On the other hand, 
the SM-like Higgs boson mass is enhanced through radiative
corrections, rendering it more easily detectable at the LHC.
Finally, the observation of a SM-like 
Higgs in the $h \to \tau \tau$ channel and not in the $h \to \gamma \gamma$ or 
vice versa, may be used to obtain additional information on the values
of the supersymmetry breaking parameters.

\vspace{1.5cm}
~\\
\large{\textbf{Acknowledgements:}} \normalsize
M.C. and C.W. would like to thank the Aspen Center for Physics, where part of 
this work was done. We wish to thank Patricia Ball, 
Thomas Becher,  Avto Kharchilava,  
Enrico Lunghi, Matthias Nuebert and Frederic Teubert.  
Work at ANL is supported in part by the US DOE, Div.\ of HEP, 
Contract DE-AC02-06CH11357. 
Fermilab is operated by Universities Research Association Inc. under contract 
no. DE-AC02-76CH02000 with the DOE. This work was also supported in part by the
U.S. Department of Energy through Grant No. DE-FG02-90ER40560.

\end{document}